\documentclass[12pt,preprint]{aastex}

\usepackage{natbib}
\newcommand\MSUN{\rm M_{\odot}}
 
\newcommand\MSUNYR{\rm M_{\odot}\,yr^{-1}}

\newcommand\MBH{M}
\newcommand\Trat{{T_{\rm i}/T_{\rm e}}}
\newcommand\grm{{\tt grmonty}}
\newcommand\ibothros{{\tt ibothros}}
\newcommand\harm{{\tt harm}}
\newcommand\GHz{{\rm\,GHz}}
\newcommand\Hz{{\rm\,Hz}}
\newcommand\Jy{{\rm\,Jy}}
\newcommand\Rg{{\,G M/c^2}}
\newcommand\sgra{{Sgr~A*}}
\newcommand\<{{\langle}}
\renewcommand\>{{\rangle}}
\newcommand\munit{{\mathcal M}}
\newcommand\lunit{{\mathcal L}}
\newcommand\tunit{{\mathcal T}}
\newcommand\mdot{{\dot{M}}}
\newcommand\muas{{\,\mu{\rm as}}}

\newcommand\ergps{{\rm erg s^{-1}}}
\newcommand\mm{{\rm\,mm}}
\newcommand\cm{{\rm\,cm}}
\newcommand\mum{{\,\mu{\rm m}}}
\newcommand\au{{\rm\,AU}}
\newcommand\hr{{\rm\,hr}}
\renewcommand\deg{{\rm\,deg}}

\newcommand\keV{{\rm\,keV}}
\newcommand\kpc{{\rm\,kpc}}
\newcommand\risco{r_{\rm ISCO}}

\slugcomment{Accepted by ApJ}

\shorttitle{Models of Sgr A*}
\shortauthors{Mo{\'s}cibrodzka et al.}

\begin{document}

\title{Radiative Models of Sgr A* from GRMHD Simulations}

\author{Monika Mo{\'s}cibrodzka\altaffilmark{1}, Charles
  F. Gammie\altaffilmark{1,2}, Joshua C. Dolence\altaffilmark{2}, \\ 
  Hotaka Shiokawa\altaffilmark{2}, Po Kin Leung\altaffilmark{2}}

\affil{$^1$ Department of Physics, University of Illinois, 1110 West Green
  Street, Urbana, IL 61801}
\affil{$^2$ Astronomy Department, University of Illinois, 1002 West Green
  Street, Urbana, IL 61801}
\email{mmosc@illinois.edu}

\begin{abstract}

Using flow models based on axisymmetric general relativistic
magnetohydrodynamics (GRMHD) simulations, we construct radiative models
for \sgra.  Spectral energy distributions that include the effects of
thermal synchrotron emission and absorption, and Compton scattering, are
calculated using a Monte Carlo technique.  Images are calculated using a
ray-tracing scheme.  All models are scaled so that the $230 \GHz$ flux
density is $3.4\Jy$.  The key model parameters are the dimensionless
black hole spin $a_*$, the inclination $i$, and the ion-to-electron
temperature ratio $\Trat$.  We find that: (1) models with $\Trat = 1$
are inconsistent with the observed submillimeter spectral slope; (2) the
X-ray flux is a strongly increasing function of $a_*$; (3) the X-ray
flux is a strongly increasing function of $i$; (4) $230 \GHz$ image size
is a complicated function of $i$, $a_*$, and $\Trat$, but the $\Trat =
10$ models are generally large and at most marginally consistent with
the $230 \GHz$ VLBI data; (5) for models with $\Trat = 10$ and $i =
85\deg$ the event horizon is cloaked behind a synchrotron photosphere at
$230 \GHz$ and will not be seen by VLBI, but these models overproduce
NIR and X-ray flux; (6) in all models whose SEDs are consistent with
observations the event horizon is uncloaked at $230 \GHz$;  (7) the
models that are most consistent with the observations have $a_* \sim
0.9$. We finish with a discussion of the limitations of our model and
prospects for future improvements.

\end{abstract}

\section{Introduction}\label{sec:1} 

Long term studies of the stellar dynamics in the central parsec of our
Galaxy indicate that the object in the center of the Milky Way is
massive and compact and is therefore likely to be a supermassive black
hole (we will use \sgra\ to refer to the radio source, the putative
black hole, and the surrounding accretion flow).  Recent estimates of
\sgra's mass $M=4.5 \pm 0.4 \times 10^6 \MSUN$ and distance $D = 8.4 \pm
0.4$ kpc (\citealt{ghez:2008}, \citealt{gillessen:2008}) indicate that
it has the largest angular size of any known black hole ($G M/(c^2
D) \simeq 5.3 {\rm \muas}$). 

\sgra\ is frequently monitored at all available wavelengths: in radio
since its discovery in 1974 \citep{balick:1974}, and more recently in
submillimeter, near-infrared (NIR), and X-rays.  It is heavily obscured in the
optical and UV ($A_V \simeq 30$ mag).  \sgra\ is a ``quiescent''
galactic nucleus because its bolometric luminosity in units of the
Eddington luminosity is low, $L_{bol} \simeq 10^{-9}L_{Edd}$. The
discovery of polarized emission at $\lambda=1.3$ mm allowed the use of
Faraday rotation to place a model dependent limit on the mass accretion
rate $2 \times 10^{-7} < \mdot < 2 \times 10^{-9} \MSUNYR$ at $r <
20\Rg$ (\citealt{bower:2005}, \citealt{marrone:2006a}).  Submillimeter VLBI
of \sgra\ shows structure at very small angular scales
\citep{doeleman:2008}.

\sgra's spectral energy distribution (SED) can be fit with 
semi-analytic quasi-spherical radiatively inefficient accretion flow
(RIAF) models (e.g. \citealt{narayan:1998}), RIAF +  outflow models
(\citealt{yuan:2003}), and with time-dependent MHD models (e.g.
\citealt{goldston:2005}, \citealt{ohsuga:2005},
\citealt{moscibrodzka:2007}).  Other workers have modeled the VLBI and
submillimeter emission (\citealt{broderick:2005}, \citealt{broderick:2006a},
\citealt{broderick:2006b}, \citealt{huang:2007}, \citealt{miyoshi:2008},
\citealt{broderick:2008}, \citealt{yuan:2009}, \citealt{dexter:2009}) 
assuming a stationary
RIAF and computing emission at submillimeter wavelengths taking into account
general relativistic effects.

In this work we simultaneously model the spectral energy distribution,
including Compton scattering, and the VLBI data using a relativistically
self-consistent approach.  We assume that accretion onto \sgra\ proceeds
through a geometrically thick, optically thin, two-temperature flow that
we model using a general relativistic MHD (GRMHD) simulation.  Black
hole spin $a_*$ is self-consistently accounted for.  We also assume that
the (likely time-dependent, anisotropic, nonthermal) state of the plasma
can be described by assigning a single temperature $T_{\rm i}$ to the ions and
a possibly different temperature $T_{\rm e}$ to the electrons. Conduction is
neglected.  

The main goal of this work is to explore how $a_*$, the inclination $i$,
and the ion-to-electron temperature ratio $\Trat$ are constrained by
the data.  Our paper is organized as follows.  In \S~\ref{sec:2} we
review broadband observations of \sgra. In \S~\ref{sec:3} we outline our
technique for computing the evolution of the accretion flow and the
emergent radiation.  In \S~\ref{sec:4} we present the results of single-
and two-temperature SED computations and compare them to the observed
SED.  We summarize and discuss the model limitations in \S~\ref{sec:5}.

\section{Observations} \label{sec:2}
Sgr A* has rich observational database in radio
(\citealt{serabyn:1997},
\citealt{falcke:1998},
\citealt{zhao:2003},
\citealt{an:2005},
\citealt{marrone:2006a}), 
NIR
(\citealt{davidson:1992}, 
\citealt{herbst:1993}, \citealt{stolovy:1996}, \citealt{telesco:1996}, 
\citealt{menten:1997}, \citealt{melia:2001}, \citealt{hornstein:2002}, 
\citealt{genzel:2003}, \citealt{eckart:2006}, \citealt{schoedel:2007}),  
X-rays (\citealt{baganoff:2001}, \citealt{baganoff:2003},
\citealt{goldwurm:2003}, \citealt{porquet:2003}, \citealt{belanger:2005},
\citealt{belanger:2006}, \citealt{porquet:2008}) 
and even  $\gamma$-rays 
(\citealt{aharonian:2004}, but see \citealt{aharonian:2008}).

In general the emission from \sgra\ in the radio band is rising with the
frequency. Below $\nu = 10 \GHz$ the spectral slope $\alpha$ ($F_{\nu} \sim
\nu^{\alpha}$) was found to be $\alpha \approx 0.1$ \footnote{Not to
be confused with the phenomenological viscosity $\alpha$ of
accretion disk theory. In this paper angular momentum transport is
calculated self-consistently in a GRMHD model.} (\citealt{serabyn:1997},
\citealt{falcke:1998}).  Between $10$ and $300 \GHz$ the spectral slope
changes to $\alpha \approx 0.5$ (\citealt{falcke:1998}, \citealt{an:2005}). 

\citet{marrone:2006b} reported that the spectral slope becomes flat or
declining between $230 \GHz$ ($1.3 \mm$) and $690 \GHz$ ($0.43 \mm$),
consistent with a transition from optically thick to optically thin
radiation. He estimated a variance-weighted mean value of $\alpha=-0.18$
from four observational epochs (each epoch lasting around 2 \hr, and
$\alpha$ changing from -0.46 to 0.08). The quiescent NIR counterpart of
\sgra\ has been reported by \citet{genzel:2003}, but it is not clear
whether the ``quiescent'' NIR emission at the position of \sgra\ is background
noise or a real detection of \sgra\ (\citealt{do:2009}).  Thus, the
measured quiescent emission in the NIR is usually interpreted as an
upper limit.  The quiescent luminosity at the 2-8 keV band measured with
the {\em Chandra} observatory is $L_X=2.4 \times 10^{33} \ergps$, and
the emission is extended with an intrinsic size of $1.4\arcsec$,
consistent with the Bondi radius.  The mass accretion rate at the Bondi
radius deduced from X-ray observations is $\sim 10^{-6} \MSUNYR$
\citep{baganoff:2003}.  Above the stationary emission \sgra\ exhibits
intraday variability at all observed wavelengths (flares in
submillimeter, NIR, and X-rays that often rise simultaneously).

High frequency VLBI constrains the structure of \sgra\ on angular scales
comparable to $G \MBH/(c^2 D) \simeq 5.3 {\rm \mu as}$.  The
distribution of intensities on the sky is a convolution of the
(wavelength dependent) intrinsic angular structure with anisotropic
interstellar broadening proportional to $\lambda^2$
(\citealt{bower:2006}, \citealt{doeleman:2008}).  \sgra\ has been
detected by $\lambda=1.3$ mm VLBI on baselines between Hawaii (JCMT),
Arizona (SMTO), and California (CARMA) \citep{doeleman:2008}.  This
small number of baselines does not permit imaging of the emitting region
or the ``silhouette'' of the black hole (\citealt{bardeen:1973},
\citealt{chandrasekhar:1983}, \citealt{falcke:2000},
\citealt{takahashi:2004}), but it does constrain models of the emitting
region.  Using a (two-parameter) symmetric Gaussian brightness
distribution model \cite{doeleman:2008} infer a full width at half
maximum FWHM $= 37^{+16}_{-10} {\rm \mu as}$, or $\simeq 7 G \MBH/(c^2
D)$.  This is {\em very small}, since the apparent diameter of the black
hole is $2 \sqrt{27} G \MBH/(c^2 D) \simeq 55 {\rm \mu as}$. The FWHM
for a Gaussian model has also been estimated at $7 \mm$ ($0.268
\pm{0.025}$ mas, or $51 G \MBH/(c^2 D)$, \citealt{bower:2004}) and at
$3.5 \mm$ ($0.126 \pm{0.017}$ mas, or $28 G \MBH/(c^2 D)$,
\citealt{shen:2005}), but longer wavelength intrinsic size is more
difficult to measure because scatter broadening dominates the observed
image size at $\lambda \gtrsim 1 \mm$.  VLBI observations at
$\lambda=0.8 \mm$ ($345 \GHz$) and $0.65 \mm$ ($450 \GHz$) are expected
in the near future \citep{doeleman:2009}.

\section{Model} \label{sec:3}

Our model consists of three parts: a physical model of the accreting
plasma; a numerical realization of the physical model; and a procedure
for calculating the emergent radiation from the accreting plasma.

The physical model is a geometrically thick, optically thin, turbulent
plasma accreting onto a rotating black hole in a statistically steady
state. The angular momentum of the hole is assumed to be aligned with
the angular momentum of the accreting plasma. \footnote{Tilted, or
``oblique'' accretion flows, require 3D simulations; antialigned flows
can be modeled using an axisymmetric simulation, but likely provide a
worse fit to the data than the low-spin aligned flows considered here.}
The ions and electrons are assumed to have a thermal distribution
function, but with a temperature ratio $\Trat$ that may be different
from one \citep[see  \S 2.2.1.~in][for a discussion of temperature
ratios in a collisionless accretion flow model; their work suggests that
$\Trat \sim 10$ may be a natural value]{sharma:2007}.  The equation of
state is gas pressure $P = (\gamma - 1) u$ ($u$ is the proper internal
energy density), with $\gamma = 13/9$, appropriate to a plasma with
$\Theta_i \equiv k T_{\rm i}/(m_p c^2) \ll 1$ and $\Theta_e \equiv k
T_{\rm e}/(m_e c^2) \gg 1$ (we will discuss our procedure for extracting
an electron temperature later).  The parameters of the accreting plasma
model, then, are $a_*$ and $\Trat$.

The numerical realization of the physical model uses the GRMHD code
\harm, a conservative shock-capturing scheme with constrained transport
to preserve $\nabla \cdot {\mathbf B} = 0$ \citep{gammie:2003}.  All models in
this paper are {\em axisymmetric}; we will explore 3D models in a
subsequent publication.  Our grid is uniform in modified spherical
Kerr-Schild coordinates \citep{gammie:2003}, which permit the flow to
be followed through the event horizon.  The coordinates are logarithmic
in the Kerr-Schild radius $R$ and nonuniform in Kerr-Schild colatitude
$\theta$ (Boyer-Lindquist and Kerr-Schild $R$ and $\theta$ are
identical).\footnote{The modified Kerr-Schild coordinates
$x_0$,$x_1$,$x_2$,$x_3$ are related to spherical Kerr-Schild coordinates
$t, R, \theta, \phi$ by $t = x_0$, $R = e^{x_1}$, $\theta=\pi x_2 +
\frac{1}{2} (1-h) \sin(2 \pi x_2)$, and $\phi = x_3$.  We set $h = 0.3$.}
The resolution is $256 \times 256$.  

The inner boundary of the computational domain is at $R_{in} = 0.98 (1 +
\sqrt{1 - a_*^2})$, i.e.  just inside the event horizon.  The outer
boundary is at $R_{\rm out} = 40 \Rg \simeq 1.8\au$, or an angular radius of
$\simeq 210 {\rm \muas}$.  Since low frequency emission is believed to arise at
larger radius, this means that we are unable to model the low frequency
(radio and mm) portion of the SED.  

We must also supply initial conditions and boundary conditions.  For
numerical convenience we adopt the same initial equilibrium torus used
by \citealt{gammie:2003}, \citealt{mckinney:2004}, and others.  This
torus has an inner boundary at $6\Rg$ and a rest-mass density maximum
$\rho_{\rm max}$ at $R_{\rm max} = 12\Rg$.\footnote{For $a_* = 0.5$ we
set $R_{\rm max} = 13\Rg$ so that scale height $H$ has $H/R_{\rm max}$
similar for all the models.} It is seeded with a weak, purely poloidal
field that follows the isodensity contours and has minimum $\beta \equiv
P/(B^2/(8\pi)) = 100$ (see \cite{gammie:2003} for details).  Small
perturbations are added to the internal energy.  The torus quickly
becomes turbulent due to the magnetorotational instability
\citep{balbus:1991}.  At $R \lesssim R_{\rm max}$ the accretion flow
soon reaches a nearly (statistically) stationary state that is
independent of the initial conditions (except for the magnetic field
geometry; see \citealt{hawley:2002}, \citealt{beckwith:2008}).  If our
numerical model accurately represents the physical model, this inner
accretion flow should be similar to the inner portion of a much more
extended accretion flow.  We use outflow boundary conditions at both the
inner and outer boundaries, and the usual polar boundary conditions at
$\theta = 0$ and $\theta = \pi$.  We integrate for $2000 G \MBH/c^3
\simeq 12\hr$, or 8 orbital periods at $R_{\rm max}$.

\harm\ (and similar codes) fail if $\rho$ or $u$ are small in comparison
to the kinetic and magnetic energy densities, or the density in nearby
zones.  To prevent this we impose a hard ``floor,'' so that $\rho >
10^{-4} \rho_{\rm max} (R/\Rg)^{-3/2}$ and $u > 10^{-6} \rho_{\rm max}
c^2 (R/\Rg)^{-5/2}$.  

To ``observe'' the numerical model we must specify the observer's
distance $D$ and the inclination $i$ of the black hole spin to the line
of sight.  Because the dynamical simulation is scale-free but the
radiative transfer calculation is not, we need to specify the simulation
length unit $\lunit = GM/c^2$, time unit $\tunit = G M/c^3$, and mass
unit $\munit$ (equivalently: the mass accretion rate).  Since we set the
peak density in the GRMHD model to $1$ in simulation units, the peak
density is $\munit/\lunit^3$ in cgs units.  The mass unit is {\em not}
set by $M$ (which appears in the dynamical model only in the combination
$GM$)  because the flow mass is $\ll M$, and has negligible effect on
the gravitational field.   $\munit$ is therefore a free parameter.

To calculate the SED we use the relativistic Monte Carlo scheme \grm.  A
detailed description of the algorithm and tests are given in
\cite{dolence:2009}.  The code fully accounts for synchrotron emission
and absorption, and Compton scattering.  It uses a ``stationary flow''
approximation, computing the spectrum through each time slice of
simulation data as if it were time-independent.  This is an
approximation because the light crossing time is comparable to the
dynamical time.  It is done because tracking photons through the
time-dependent simulation data is still too computationally expensive.
We will evaluate the quality of this approximation once we are able to
calculate fully self-consistent spectra.  An average spectrum is formed
by averaging over $50$ single slice spectra from each of $4$ different
realizations of the simulation (the realizations differ in the random
number seed used to generate initial conditions).

To image the model we use the relativistic ray-tracing code \ibothros,
which accounts only for synchrotron emission and absorption, again using
a stationary flow approximation \citep[see][]{noble:2007}.  An average
image is created using the same averaging procedure as for the spectra.

To sum up, the model parameters (aside from those that describe the
initial conditions) are $M$, $D$, $i$, $a_*$, $\Trat$, and $\munit$.
$M$ and $D$ are set by the observations of stellar orbits; we briefly
explore the consequence of varying them below. $\munit$ we will set for
each model by requiring that time-averaged flux $F_\nu(230\GHz) =
3.4\Jy$.  The remaining three free parameters are $a_*$, $i$, and
$\Trat$.

In what follows we explore models with $a_* = 1 - 2^{-n}$, with $n =
1,2,3,4,5$ and $6$, i.e. $a_* \approx 0.5, 0.75, 0.88, 0.94, 0.97$ and $0.98$,
$\Trat = 1, 3$, and $10$, and $i = 5, 45$ and $85\deg$.

The observational constraints on the model are (aside from the $230\GHz$
flux density): the submillimeter spectral slope $\alpha$, the upper
limit on the quiescent NIR flux density (\citealt{genzel:2003},
\citealt{melia:2001}, \citealt{hornstein:2002}, \citealt{schoedel:2007},
\citealt{dodds:2009}) and the upper limit on the quiescent luminosity
between $2$ and $8$ keV \citep{baganoff:2003}. Since the measured $L_X$
is produced inside $1.4\arcsec$ ($\approx 10^5 \Rg$), and our simulation
domain covers only the inner $40 \Rg$, we exclude models for which $L_X$ is
close to the ``quiescent emission'' $L_X$. \footnote{We use the source
brightness profile \citep{baganoff:2003} and estimate that $50\%$ of the
X-ray luminosity comes from the central pixel of size $0.492\arcsec$.
This is still far larger than our computational domain, so we require
that the X-ray luminosity in our models not exceed $L_X = 1.2 \times
10^{33} \ergps$.}  

\section{Results}\label{sec:4}

We have studied the combinations of model parameters listed in
Table~\ref{tab:1} ($\Trat$=1), Table~\ref{tab:2} ($\Trat$=3), and
Table~\ref{tab:3} ($\Trat=$10).  The model that best satisfies the
observational constraints has $a_* = 0.94$ (model D), $\Trat = 3$, and
$i = 85\deg$. We will present this ``best-bet'' model in some detail
before going on to the full parameter survey (see \S\ref{sec:4.2}) to
give the reader a physical sense for the models.

\subsection{Best-bet model}\label{sec:4.1}

\begin{figure*}
\begin{picture}(0,550)

\put(-160,480){\includegraphics{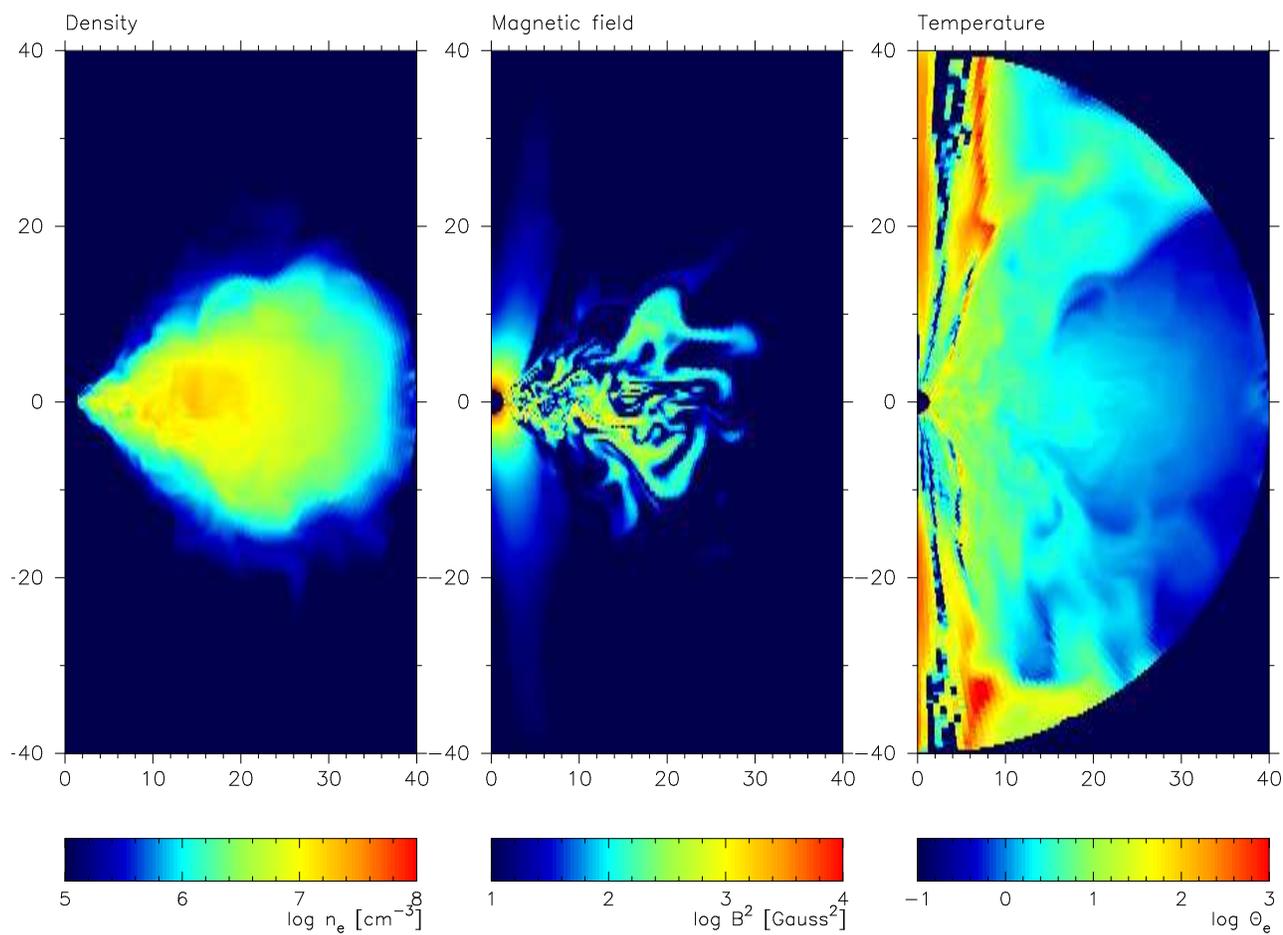}}

\end{picture}
\caption{Disk structure in our best-bet model with $a_* = 0.94$
(model D) and with $\Trat=3$. The number density, the magnetic field
strength, and the electron temperature are
shown in the left, middle, and right panel respectively. The axis scale
units are $\Rg$ .The figure
presents a single time slice.}
\label{fig:physvar}
\end{figure*}

In what follows we will consider time and realization averaged SEDs and
images.  But to get a sense for physical conditions in the accretion
flow,  consider a single time-slice at $t = 1680 \, GM/c^3$.
Figure~\ref{fig:physvar} shows the run of $n_e, B^2$, and $\Theta_e$ in
the time slice.  Evidently $10^7 \cm^{-3}$ is a typical equatorial
plane density, $30$G is a typical field strength, and $\Theta_e = 10$ is
a typical electron temperature.  Notice that the field changes from a tangled,
turbulent structure near the midplane to a more organized structure in
the ``funnel'' over the poles of the black hole.  Temperature generally
increases away from the midplane.

\begin{figure*}
\begin{picture}(0,500)
\put(-120,-50){\includegraphics{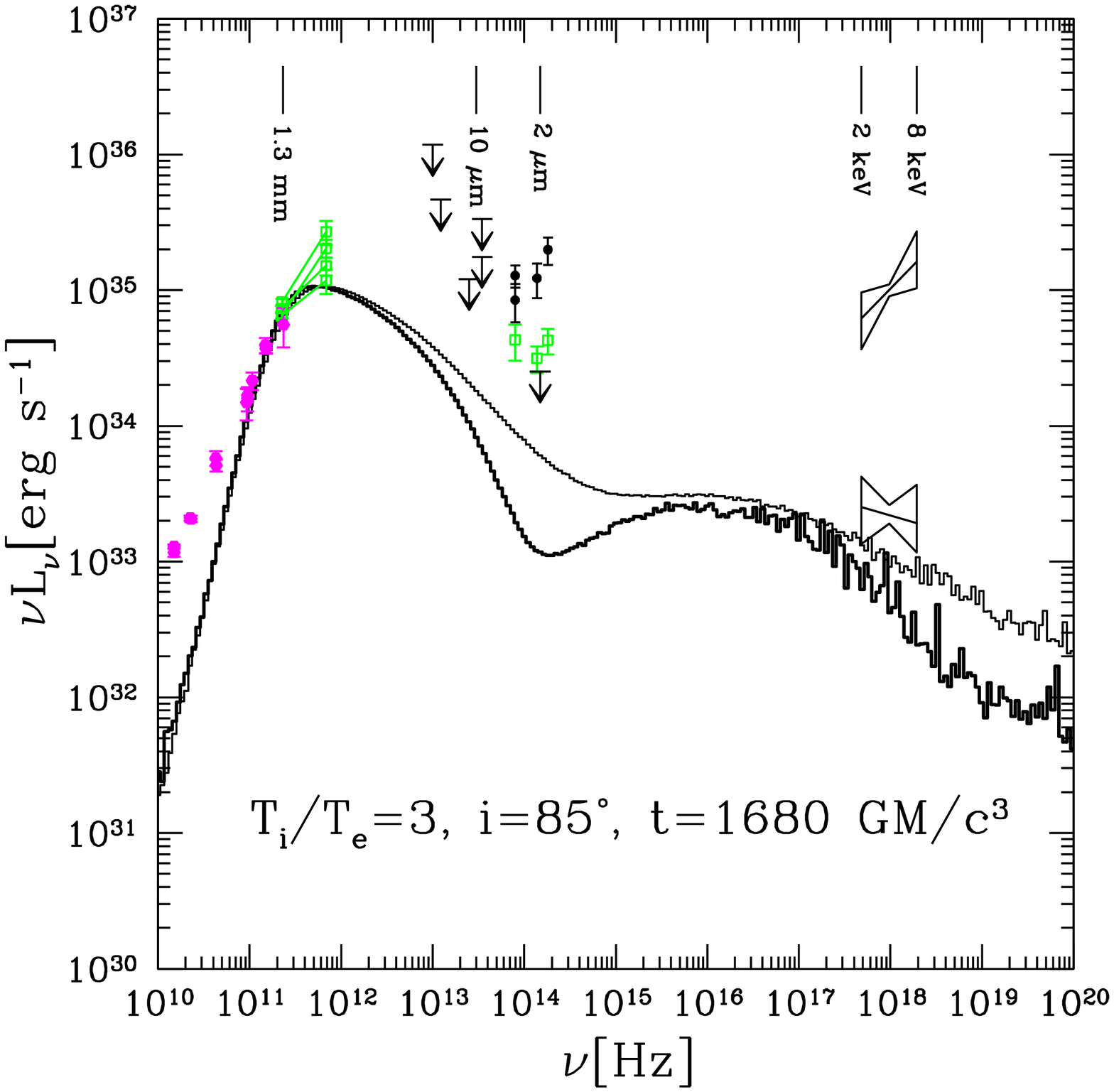}}
\end{picture}
\caption{SEDs computed based on a single time slice $t=1680 G M/c^3$ 
(thick line) (see
Figure~\ref{fig:physvar} for the distributions of the physical variables
corresponding to the same time) along with the time averaged spectrum (thin
line) of our best-bet model.
Observational points are taken from:
\citealt{falcke:1998}, \citealt{an:2005}, \citealt{marrone:2006a} at radio,
\citealt{genzel:2003} at NIR (1.65, 2.16, and 3.76 ${\rm \mu m}$) and
\citealt{baganoff:2003} at X-rays (2-8 keV). The upper limits in the NIR band
are taken from \citealt{melia:2001} (30, 24.5 and 8.6 ${\rm \mu m}$),
\citealt{schoedel:2007} (8.6 $\mu m$) and \citealt{hornstein:2007} (2 ${\rm \mu
m}$).  The points in the NIR at flaring state are from \citealt{genzel:2003}
(1.65, 2.16, and 3.76 ${\rm \mu m}$) , and \citealt{dodds:2009} (3.8 ${\rm \mu m}$).  An
example of X-ray flare ($L_X = 1 \times 10^{35}$ $\ergps$) is taken
from \citealt{baganoff:2001}.}
\label{fig:fidspect}
\end{figure*}

In Figure~\ref{fig:fidspect} we show the SED computed from the same
timeslice used in Figure~\ref{fig:physvar} (thick line) along with the
average SED (thin line).  The figure also shows a selection of radio,
NIR and X-ray observational data points (references given in the figure
caption).  The SED has a peak at $\sim 690 \GHz$ due to thermal
synchrotron emission.  Below $\sim 100\GHz$ it fails to fit the data
because that emission is produced outside the simulation volume.  A
second peak in the far UV is due to the first Compton scattering order,
and at $\gtrsim 10^{19} {\rm Hz}$ ($\gtrsim 40\keV$) the photons are produced
by two or more scatterings.

\begin{figure*}
\begin{picture}(0,550)

\put(-160,480){\includegraphics{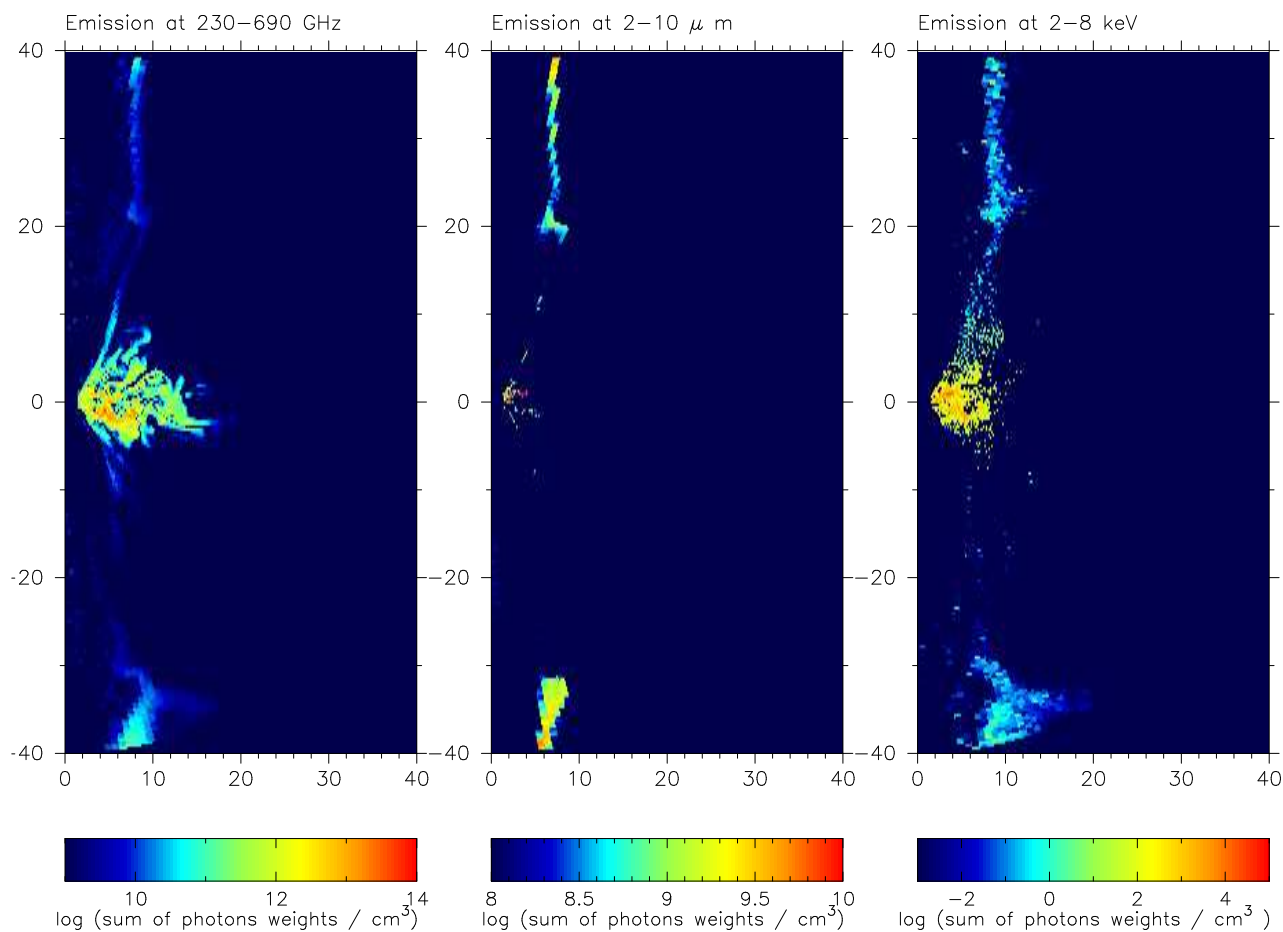}}

\end{picture}
\caption{Maps showing the point of origin for photons in our
best-bet model.  We show the logarithm of the sum of the photon
weights in each zone, which is proportional to the number of photons 
seen at $230-690 \GHz$
(left), $2-10 {\rm \mu m}$ (middle) and 2-8 keV (right) band. 
The axis scale units are $\Rg$.
The figure presents a
single time slice at $t=1680 G M/c^3$.  Note that color bands differ in scales.}
\label{fig:emmap}
\end{figure*}

Where do the photons originate in each band? Figure~\ref{fig:emmap} maps
the points of origin for photons in the synchrotron peak ($230$ to $690
\GHz$), in the NIR ($2-10\mum$) and in the X-ray (2-8 keV).  Most of the
submillimeter emission originates near the midplane at $4 \Rg < r < 6
\Rg$.  NIR photons are produced in hot regions with high magnetic field
intensity and high gas temperature, and these are concentrated close to
the innermost stable circular orbit (ISCO; $r_{ISCO} \approx 2 \Rg$ for
$a_* = 0.94$) i.e. they come from between $2 \Rg < r < 6 \Rg$.  All
photons responsible for the formation of the first Compton bump are
up-scattered between $2 \Rg < r < 8 \Rg$ but the 2-8 keV emission is
produced mainly by scatterings in the hottest parts of the disk at  $2
\Rg < r < 3 \Rg$.  Emission can be seen around the borders of the funnel
in each panel, but this is at a low level and is associated with
unreliable temperatures assigned by \harm\ to the funnel region.

The photons that form the submillimeter peak ($690 \GHz$) originate at
$\<r\> \approx 4 \Rg$, where $\<n_e\> \approx 4 \times 10^6 \cm^{-3}$
$\<B\> \approx 28$ Gauss, and $\<\Theta_e\> \approx 17$. \footnote{
Angle brackets indicate an average over grid zones weighted with the
photon number `detected' in a given frequency band.} The emissivity $\nu
j_\nu(\theta=\pi/2)$ calculated from these averaged values peaks at $\nu
\sim 8 \times 10^{11}$ Hz, which is quite close to the actual peak in
$\nu L_\nu$ (where $\nu L_\nu = 4 \pi D^2 \nu F_\nu$). The averaged
values also yield $B^2/(8 \pi P) \simeq 0.25$; the submillimeter peak
plasma is dominated by gas pressure rather than magnetic pressure.

The optically thin synchrotron photons in the energy range between the
synchrotron peak and first-order Compton scattering bump ($6.9 \times
10^{11} < \nu < 1.5 \times 10^{14}$ Hz) tend to arise in current sheets
at $2 \Rg < r < 6 \Rg$, where $1.5 < n_e/(10^6 \cm^{-3}) < 4$, $\<B\> =
30-35$~Gauss, and $\Theta_e \approx 80$ or higher. The corresponding
$\nu j_\nu$ peaks in the mid-IR, at $12 {\rm \mum}$.  These current
sheets have higher entropy than the surrounding plasma, consistent with
the idea that they are heated by (numerical) reconnection.  The
prominence of the current sheets is likely an artifact of axisymmetry.
Similar current sheets are seen in axisymmetric shearing box models of
disks that are not present in three dimensional shearing box
models.\footnote{Our best guess is that well resolved 3D models will
exhibit heating that is more evenly distributed.  Direct, physical
heating of the electrons can take place in this strongly shearing region
via plasma instabilities acting on anisotropic pressure, as discussed by
\citet{sharma:2007}.}

The synchrotron photons Compton scatter at $2 \Rg < r < 8 \Rg$ where $16
< \Theta_e < 70$.  The X-ray emission at 2-8 keV is formed by
scatterings from plasma with $\<\Theta_e\> = 35$ at $\<r\>=3 \Rg$. For
this temperature the average energy amplification per scattering is $1 +
4\Theta_e + 16 \Theta_e^2 \approx 1.9 \times 10^4$, consistent with the
seed photons with energies $2.5 \times 10^{13} < \nu < 10^{14} \Hz$ ($11
{\rm \mum} > \lambda >  2.9 {\rm \mum}$). This means that many of the
seed photons are produced in current sheets, and so some uncertainty
attaches to the Compton scattered flux. We know observationally that
\sgra\ produces frequent flares with fluxes larger than those produced
by our quiescent-source model, so there is a source of seed photons in
this energy band, albeit a fluctuating one.

A small fraction of photons are emitted from the funnel wall at
large radii ($15-40 \Rg$) where the gas temperature is $\Theta_e \sim
10^3$.  This is also likely an artifact of the inability of \harm\ and
similar codes to track the internal energy of a fluid when the internal
energy is much smaller than the other energy density scales.
Nevertheless, this raises the interesting question of what the electron
distribution function {\em should} be in the funnel.  High energy
electrons might be naturally generated within this tenuous plasma by
steepening of MHD waves excited by turbulence near the equatorial plane.

\subsection{Parameter survey}\label{sec:4.2}

\begin{figure*}
\begin{picture}(0,550)

\put(-170,370){\includegraphics{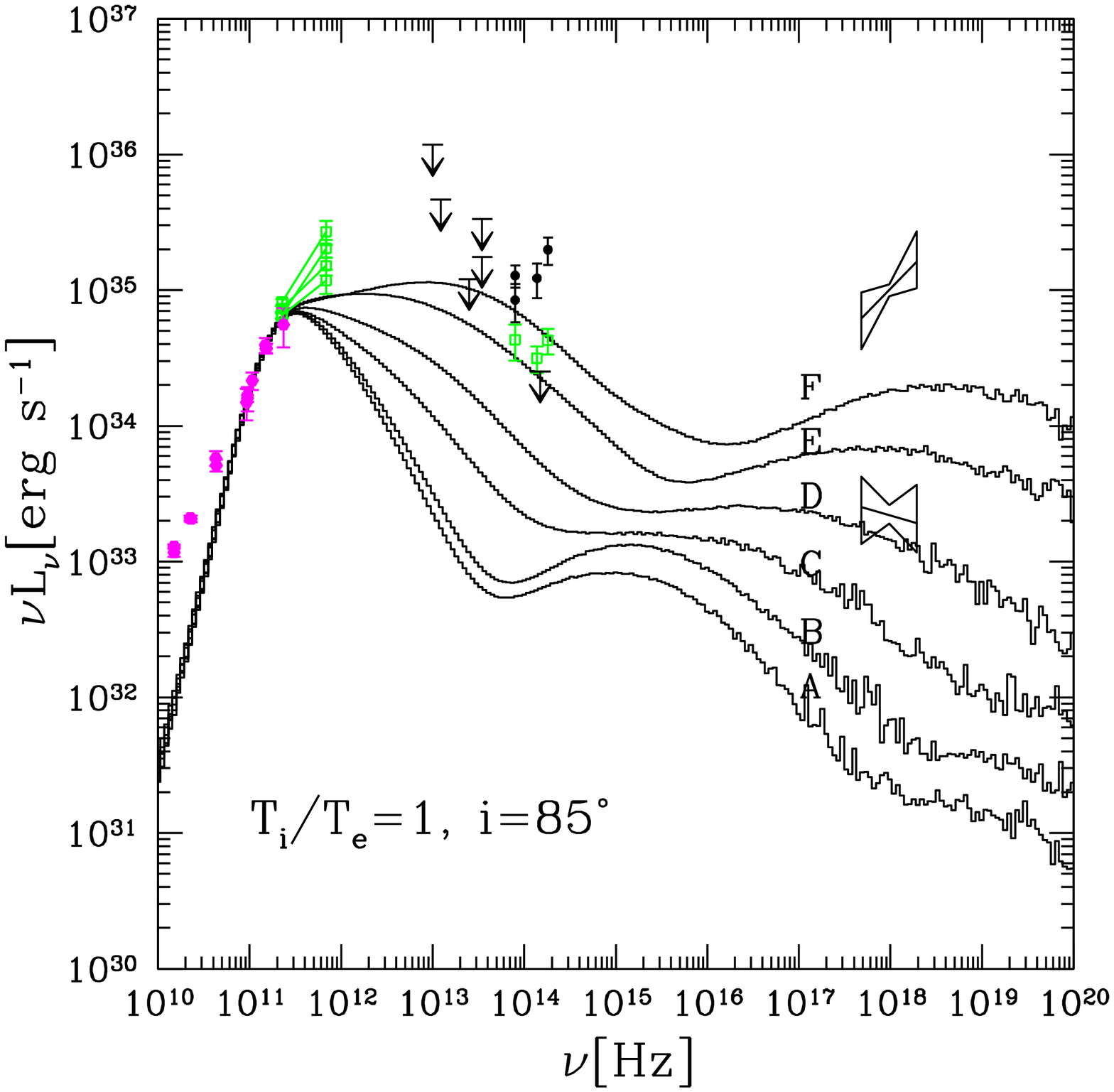}}

\put(-170,180){\includegraphics{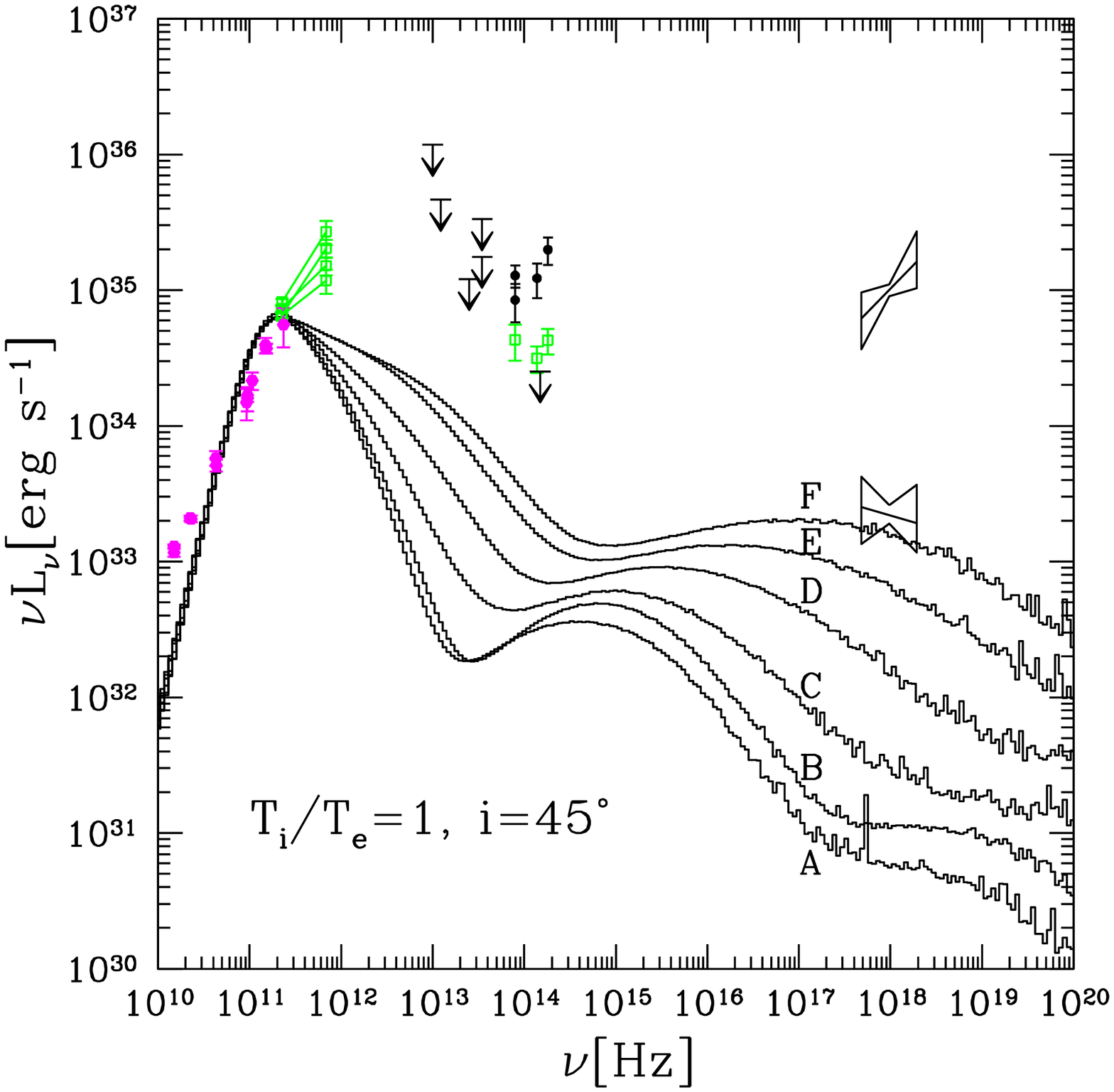}}

\put(-170,0){\includegraphics{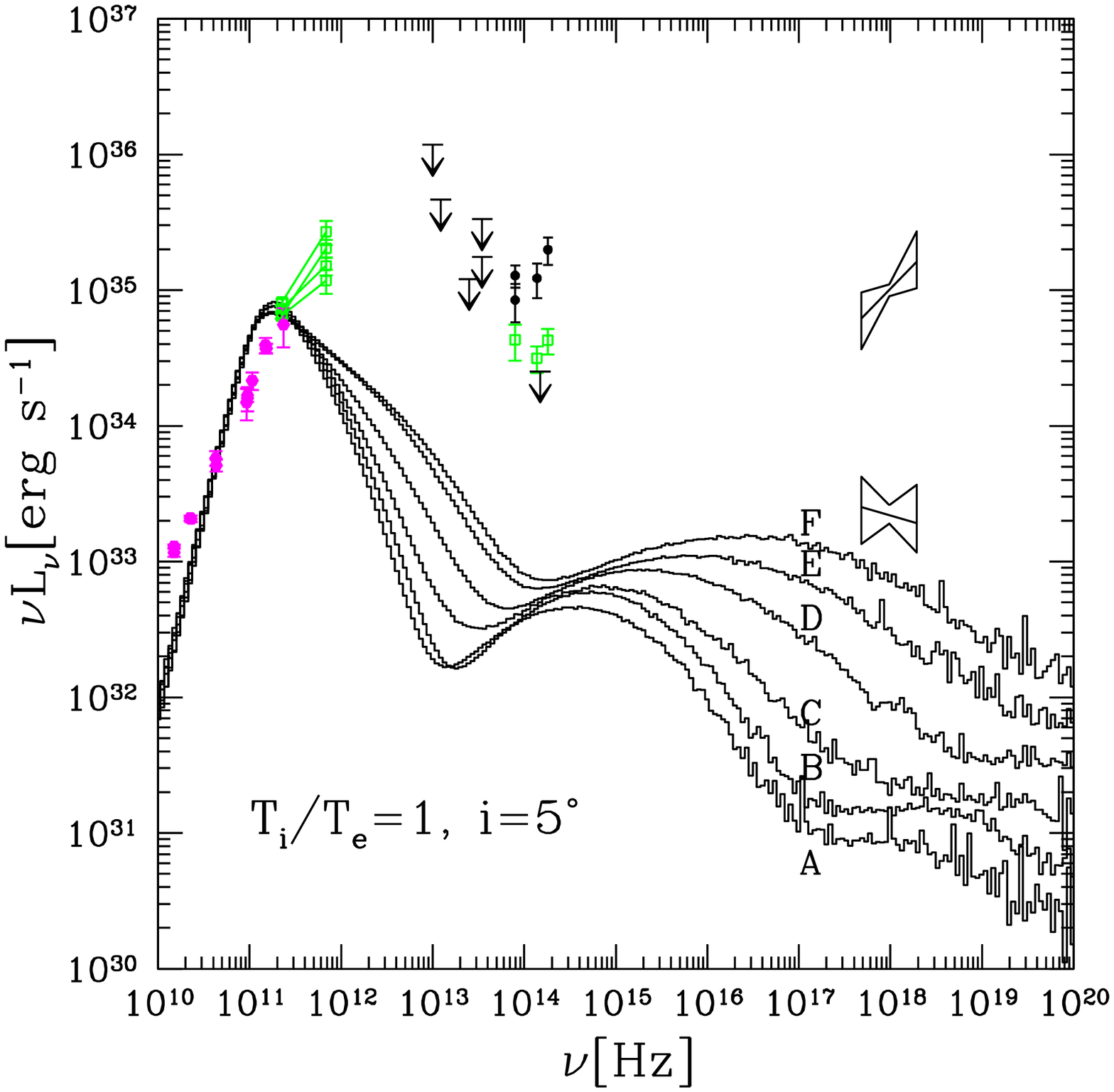}}

\put(0,370){\includegraphics{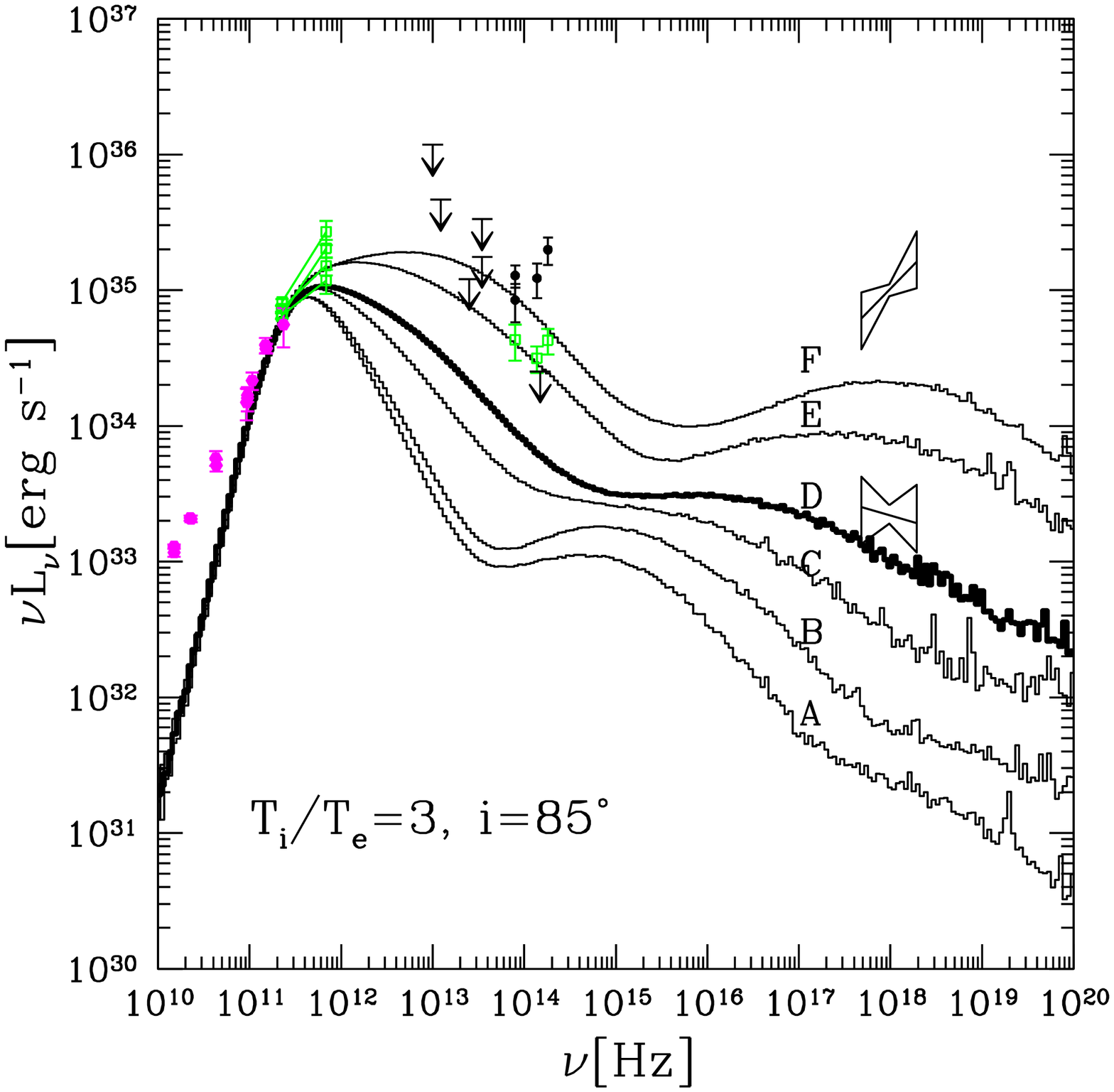}}

\put(0,180){\includegraphics{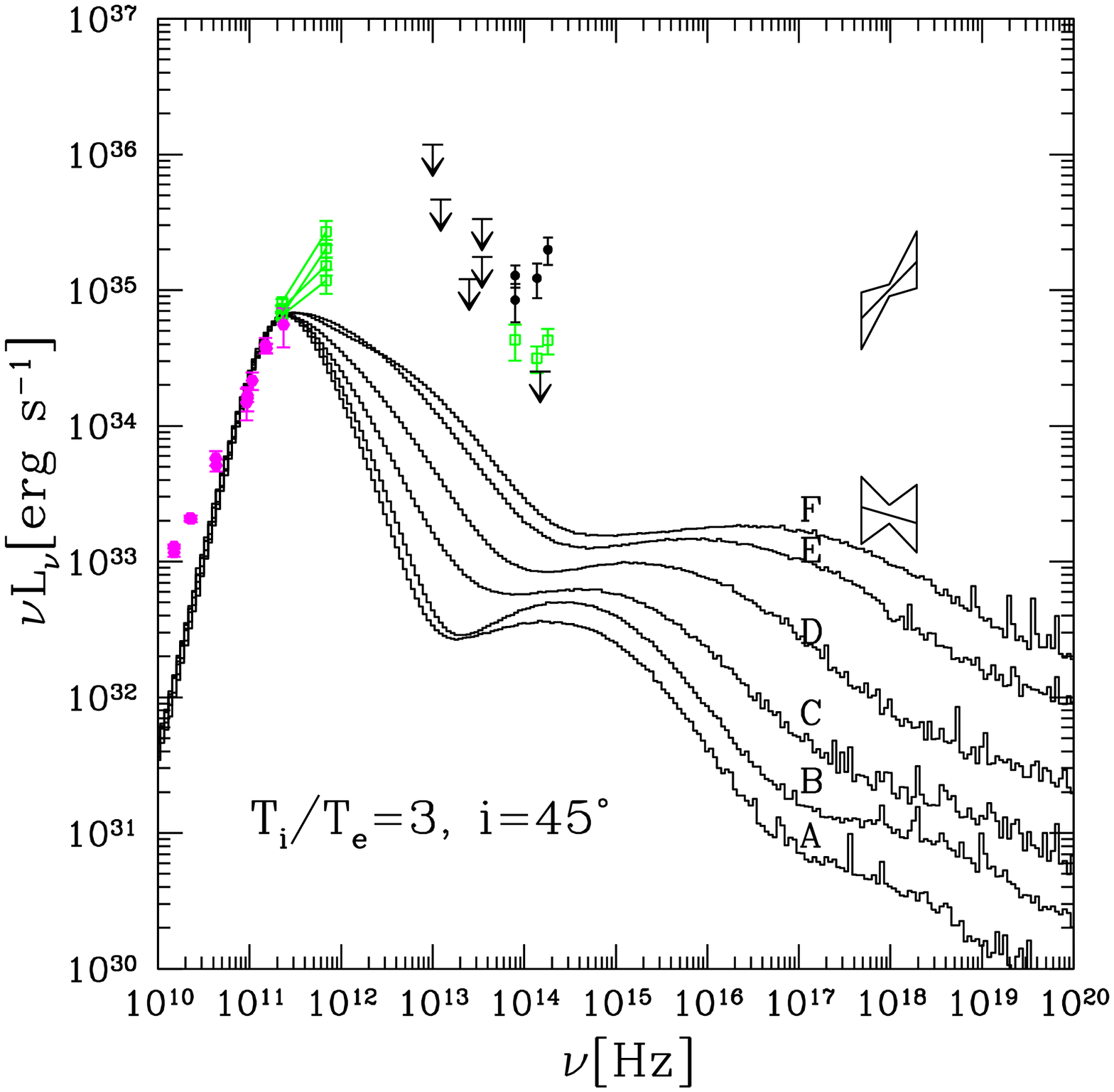}}

\put(0,0){\includegraphics{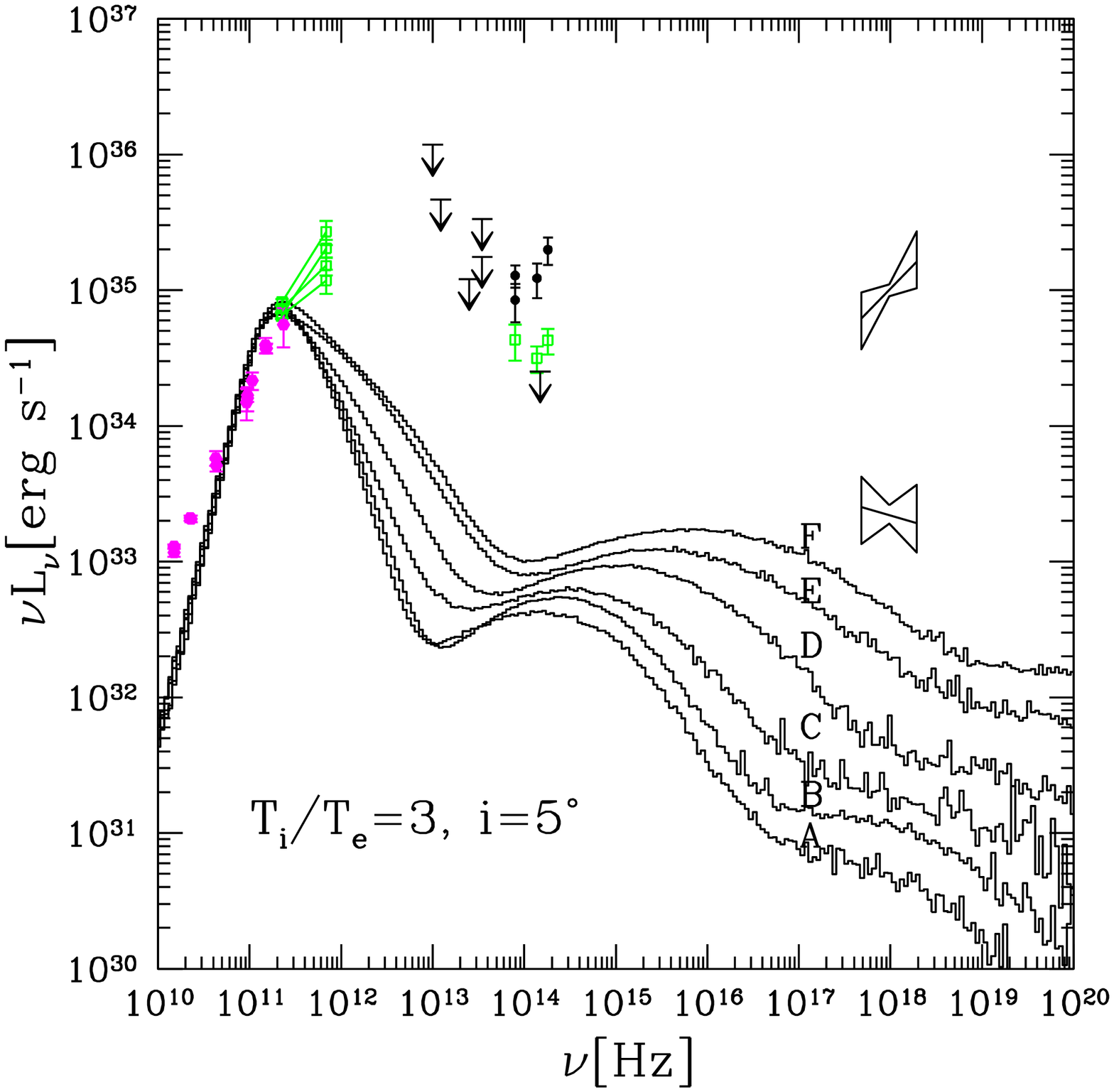}}

\put(170,370){\includegraphics{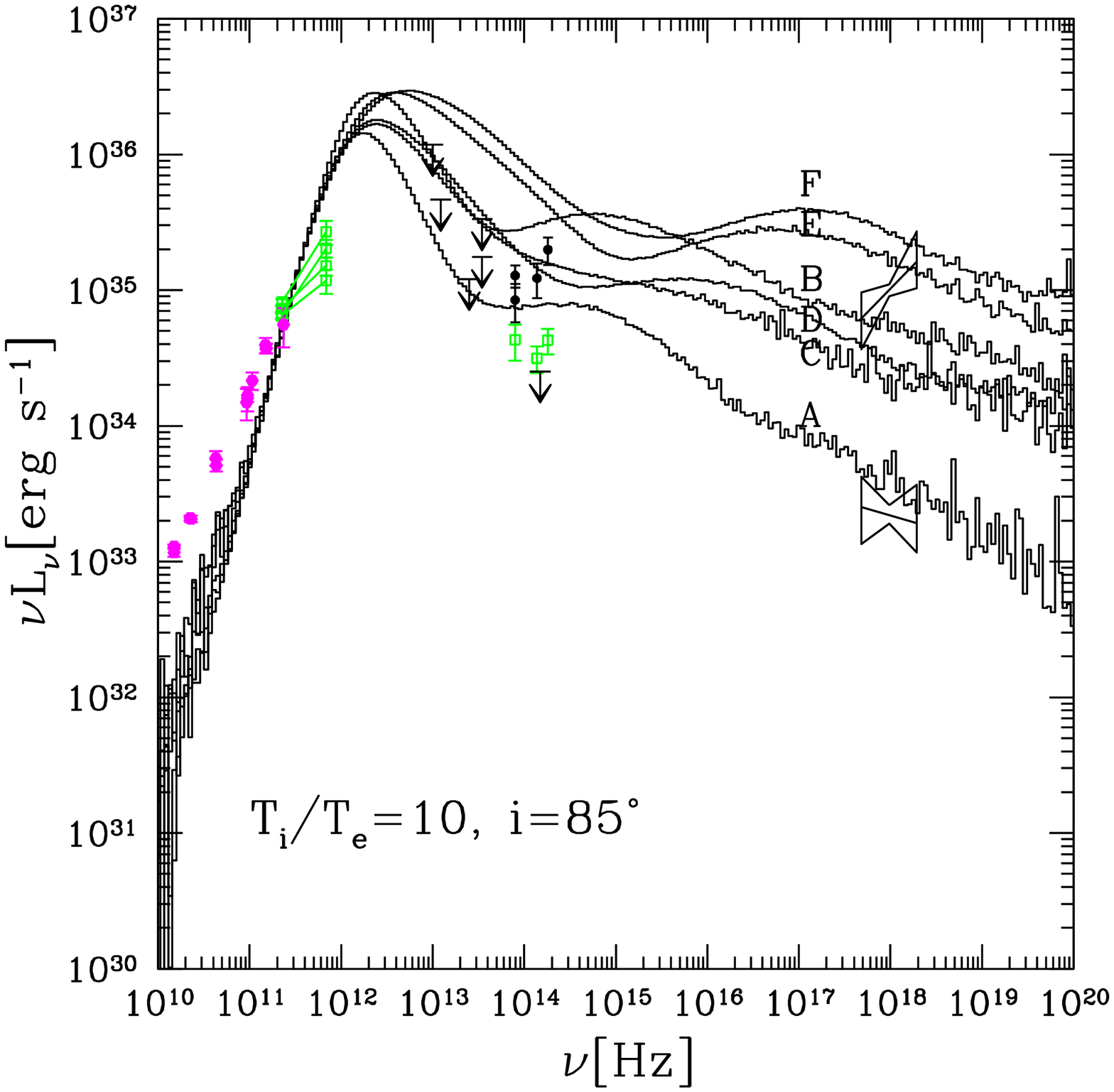}}

\put(170,180){\includegraphics{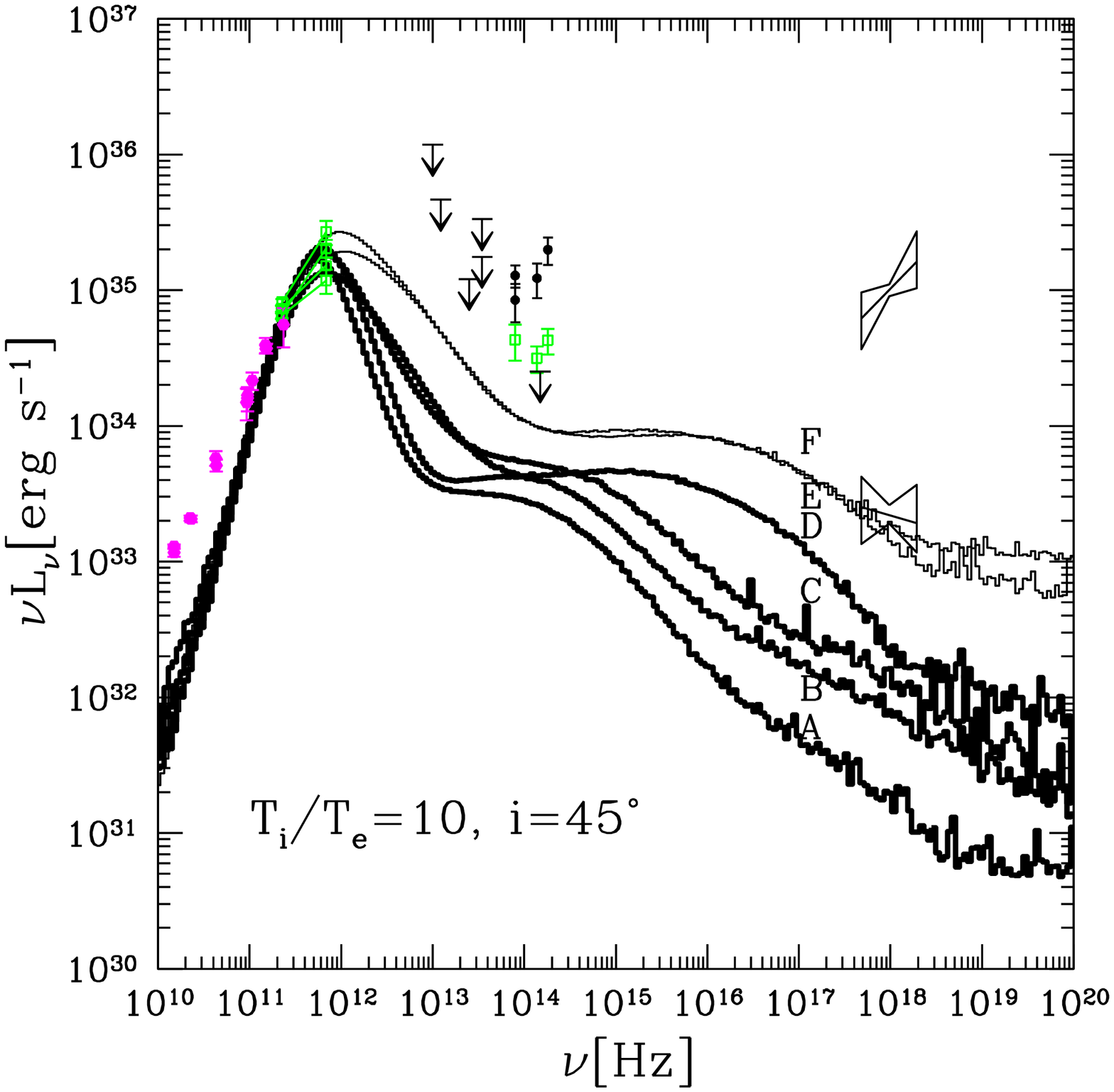}}

\put(170,0){\includegraphics{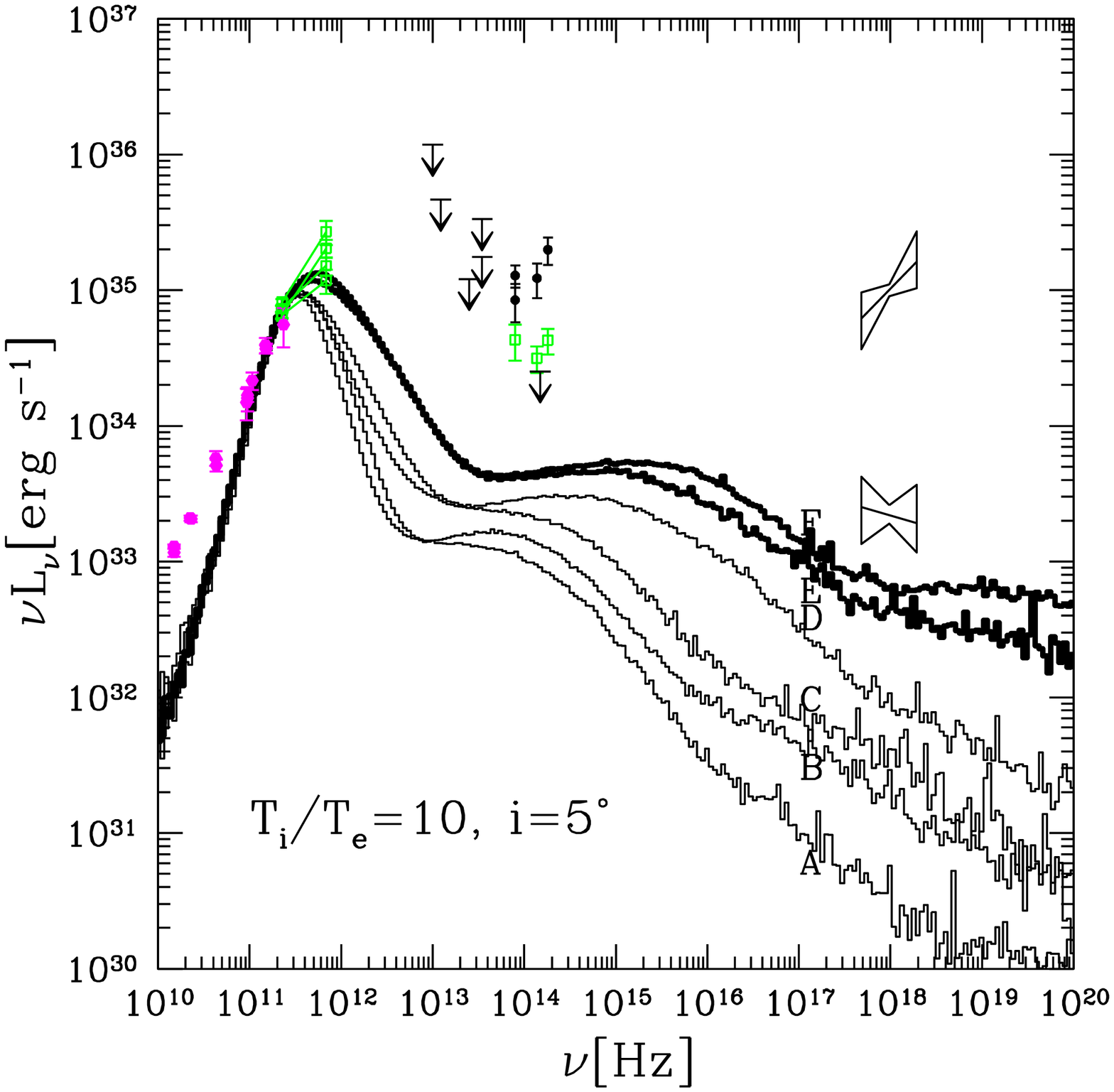}}

\end{picture}
\caption{SEDs computed for 
$\Trat=1, 3$ and 10 in left, middle and right panels, respectively and 
$i =5\deg, 45\deg$ and $85\deg$ in bottom, middle and top
panel, respectively.  Models A, B, C, D, E and F, have different $a_*$ (see
Tables~\ref{tab:1},~\ref{tab:2}, and~\ref{tab:3}).  Each SED is a result of
averaging over 200 individual SEDs taken from last $500 G M/c^3$ of the 
GRMHD runs (4
runs and 50 dumps for each spin). Observational points and upper limits as in
Figure~\ref{fig:fidspect}. Models consistent with the observations are marked
with thick lines.}
\label{fig:3}
\end{figure*}

In Figure~\ref{fig:3}, we present averaged spectra for models with
different spins (referred to as A, B, C, D, E and F; see
Tables~\ref{tab:1},~\ref{tab:2}, and~\ref{tab:3}), inclination angles
$i=85\deg, 45\deg$ and $5\deg$ in the upper, middle and bottom panels,
respectively and temperature ratio $\Trat = 1, 3,$ and $10$ from left to
right.  All SEDs are averaged over time and runs as described in
\S~\ref{sec:3}.  

The tables indicate whether the model is consistent with observations.
The model can fail in one of four ways: it can produce the wrong
submillimeter spectral slope $\alpha$; it can overproduce the quiescent
NIR flux; it can overproduce the quiescent X-ray flux; and it can be too
large at $230\GHz$ to be consistent with the VLBI data.  The last
constraint we will discuss separately in the next section.  It may be
useful to recall that $\munit$ is adjusted in each model so that the
$230\GHz$ flux is $3.4\Jy$.

The model can also fail by cooling too rapidly to be consistent with our
neglect of cooling in the dynamical model.  The Tables list a radiative
efficiency $\eta \equiv L_{\rm BOL}/\mdot c^2$, where $L_{\rm BOL}$ is
the bolometric luminosity (integrated over solid angle), and for
comparison a thin disk efficiency at the same $a_*$.  $\eta$ ranges
between $5.4 \times 10^{-4}$ for $a_* = 0.5, \Trat = 10$ to $0.18$ for
$a_* = 0.98, \Trat = 1$ (the thin disk efficiency for the latter is
$0.25$).  Only in the $a_* = 0.98$, $\Trat = 1$ model is the radiative
efficiency sufficiently high that cooling is likely to have a
significant effect on the GRMHD model.  We will consider models with
cooling in a future publication.

Very few of the time averaged SEDs based on a single-temperature ($\Trat
= 1$) models produce the correct $\alpha$.  The exception is edge-on
tori ($i=85\deg$) around fast spinning black holes (model E and F).
These models are ruled out, however, because they overproduce NIR and
X-ray flux.  

For $\Trat=3$ only the model with $a_* = 0.94$ seen at $i=85\deg$
agrees with the data.  This is the best-bet model discussed in
\S~\ref{sec:4.1}.  For $i=85\deg$, models with spins below $a_* = 0.94$
(A, B and C) are ruled out by the inconsistent spectral slope, and
models with higher spins (E and F), although consistent with the
observed $\alpha$, overproduce the quiescent NIR and X-ray emission.
All models with $\Trat=3$ observed at $i=5\deg$ and $45\deg$ are ruled
out by the inconsistent $\alpha$.

For $\Trat=10$, we find that all models with $i=85\deg$ are ruled out by
both $\alpha$ and violation of NIR and X-ray limits.  For lower
inclination angles ($i=5\deg,45\deg$) a few models (E and F with $i =
5\deg$, and A, B, C, and D at $i = 45\deg$) reproduce the observed
$\alpha$.  These models are consistent with X-rays and NIR limitations.
Models E and F for $i=45\deg$ are ruled out by NIR and X-ray limitations
whereas models A, B, C and D for $i=5\deg$ produce $\alpha$ which is too
small.

What is the physical origin of these constraints?  

The dependence on $a_*$ arises largely because as $a_*$ increases
the inner edge of the disk --- the ISCO -- reaches deeper into the
gravitational potential of the black hole, where the temperature and
magnetic field strength are higher.  In the disk mid-plane, the
temperature is a fraction of the virial temperature and scales with
radius $\Theta_e \propto 1/r$. $B \propto 1/r$, while the density $\sim
r$, below the pressure maximum. Holding all else constant (which we do
not: we hold the $230\GHz$ flux constant) this implies a higher peak
frequency for synchrotron emission, a constant Thomson depth (in our
models the Thomson depth at the ISCO is roughly constant, since the path
length $1/\sim \risco$ but the density $\sim \risco$), and a larger
energy boost per scattering $A \approx 16 \Theta_e^2$, as can be seen in
comparing models with different spin in Figure 4.  The X-ray flux
therefore increases with $a_*$ because $\Theta_e$ at the ISCO
increases.

The dependence on $\Trat$ is mainly due to synchrotron self-absorption,
which is strongest at high inclination.  For example, because the $i =
85\deg$, $\Trat = 10$ model is optically thick at $230\GHz$ the emission
is produced in a synchrotron photosphere well outside $\risco$.  The
typical radius of the synchrotron photosphere ranges between 15 $\Rg$
for low spin models ($a_*=0.5,0.75$) and 8 $\Rg$ for high spin models
($a_*> 0.75$). The $230\GHz$ flux can then be produced only with large
$\munit$; as $\munit$ increases the optically thin flux in the NIR
increases due to increasing density and field strength.   The
scattered spectrum also depends on $\Trat$ since the energy boost per
scattering is $\sim 16\Theta_e^2 \propto 1/(\Trat)^2$.

The inclination dependence is, interestingly, a relativistic effect.
$\munit$ is nearly independent of $i$ (it varies by $\sim 10\%$, except
for $\Trat = 10$, which due to optical depth effects has much larger
variation), so models with different inclination are nearly identical.
Nevertheless the X-ray flux varies dramatically with $i$, increasing by
almost 2 orders of magnitude from $i = 5\deg$ to $i = 85\deg$.  This
occurs because Compton scattered photons are beamed forward parallel to
the orbital motion of the disk gas.  The variation of mm flux with $i$
is due to self-absorption.  The mm flux reflects the temperature and
size of the synchrotron photosphere. 
At lower $i$ the visible
synchrotron photosphere is hotter than at high $i$.

There is an additional constraint due to Faraday rotation measurements,
but this constraint is qualitatively different because we do not
directly calculate Faraday rotation in our model.  Instead we adopt the
constraints on $\mdot$ which are inferred, via a separate model, from
the Faraday rotation data (\citealt{bower:2005},
\citealt{marrone:2006a}).  $\dot{M}$ increases, in a nonlinear way, with
increasing $\Trat$. For $\Trat = 1, \dot{M} > 6 \times 10^{-10}
\MSUNYR$.  For $\Trat = 10, \dot{M} < 4 \times 10^{-7} \MSUNYR$.  All
these values are consistent with the Faraday rotation constraints,
although the highest $\dot{M}$, $\Trat = 10$, models are only marginally
consistent.

There are a few other general trends worth mentioning.  In all models
the average optical depth drops below $1$ at $0.4$ to $1.3\mm$.  For
$i>45 \deg$ and high BH spins $a_* > 0.75$ the emission in NIR band (2
${\rm \mu m}$) is formed by the direct synchrotron emission while the
$2-8\keV$ emission results from a first-order scattering. For low BH
spins $a_* \le 0.75$ the emission in NIR is due to first-order Compton
scatterings and the X-ray is second-order scattering.  For $\Trat=10$
and $i=5\deg$ independently of the BH spin the NIR emission is formed by
a first-order Compton scatterings and X-rays- by second-order
scatterings.

\subsection{Images and the size of the emitting region}\label{sec:4.3}

\begin{figure*}
\begin{picture}(0,530)

\put(-120,580){\includegraphics{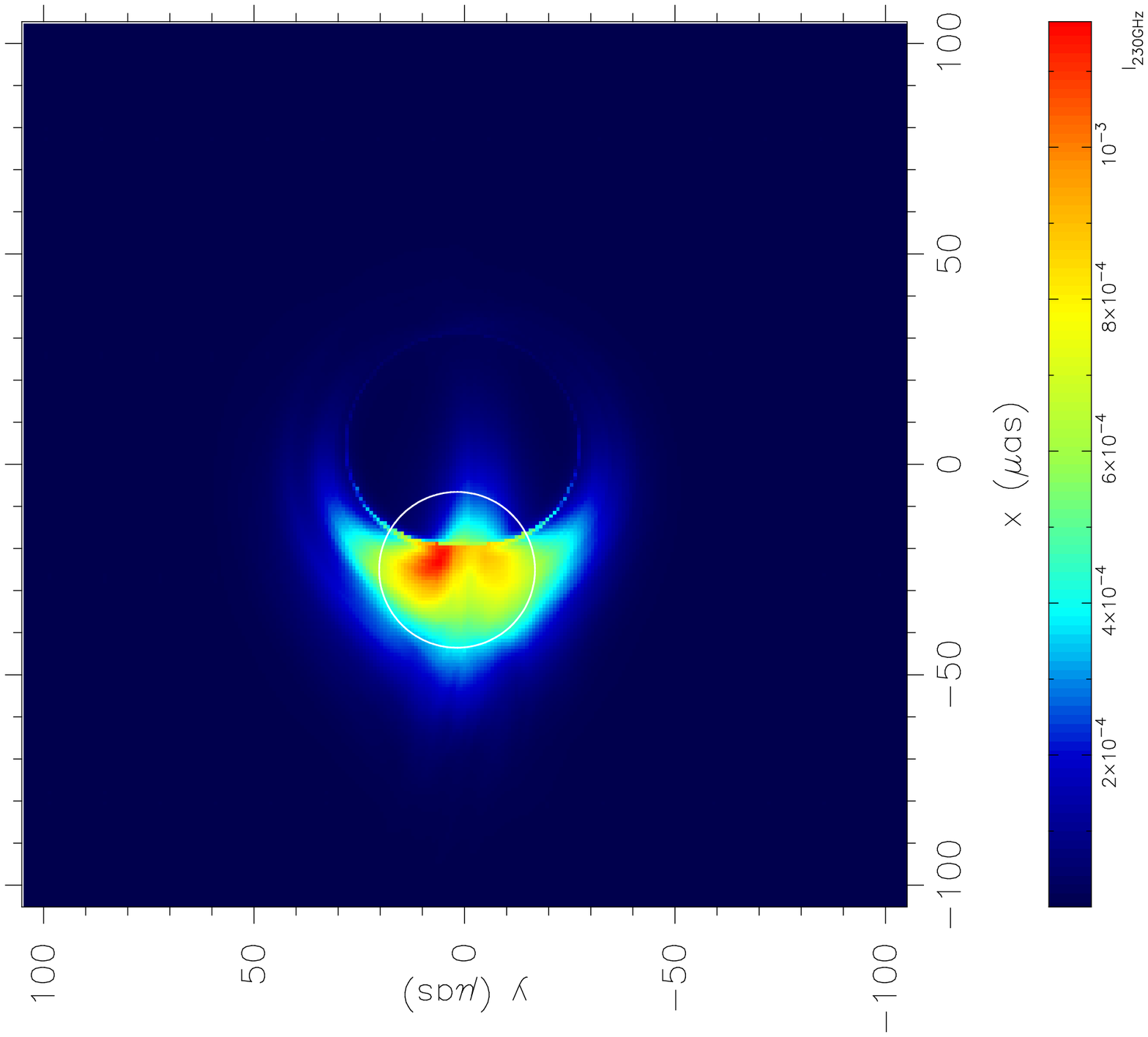}}
\put(90,580){\includegraphics{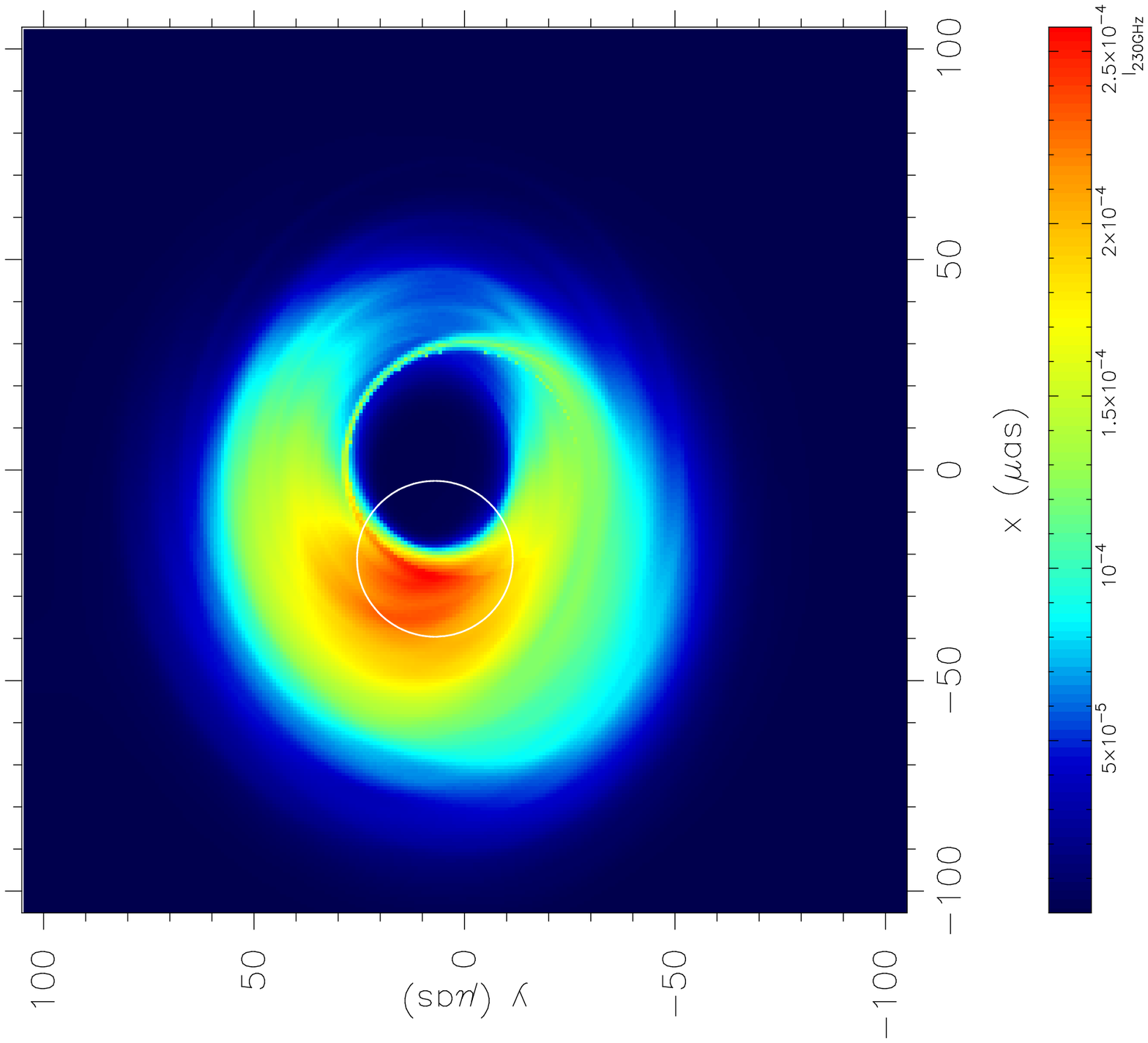}}

\put(-120,400){\includegraphics{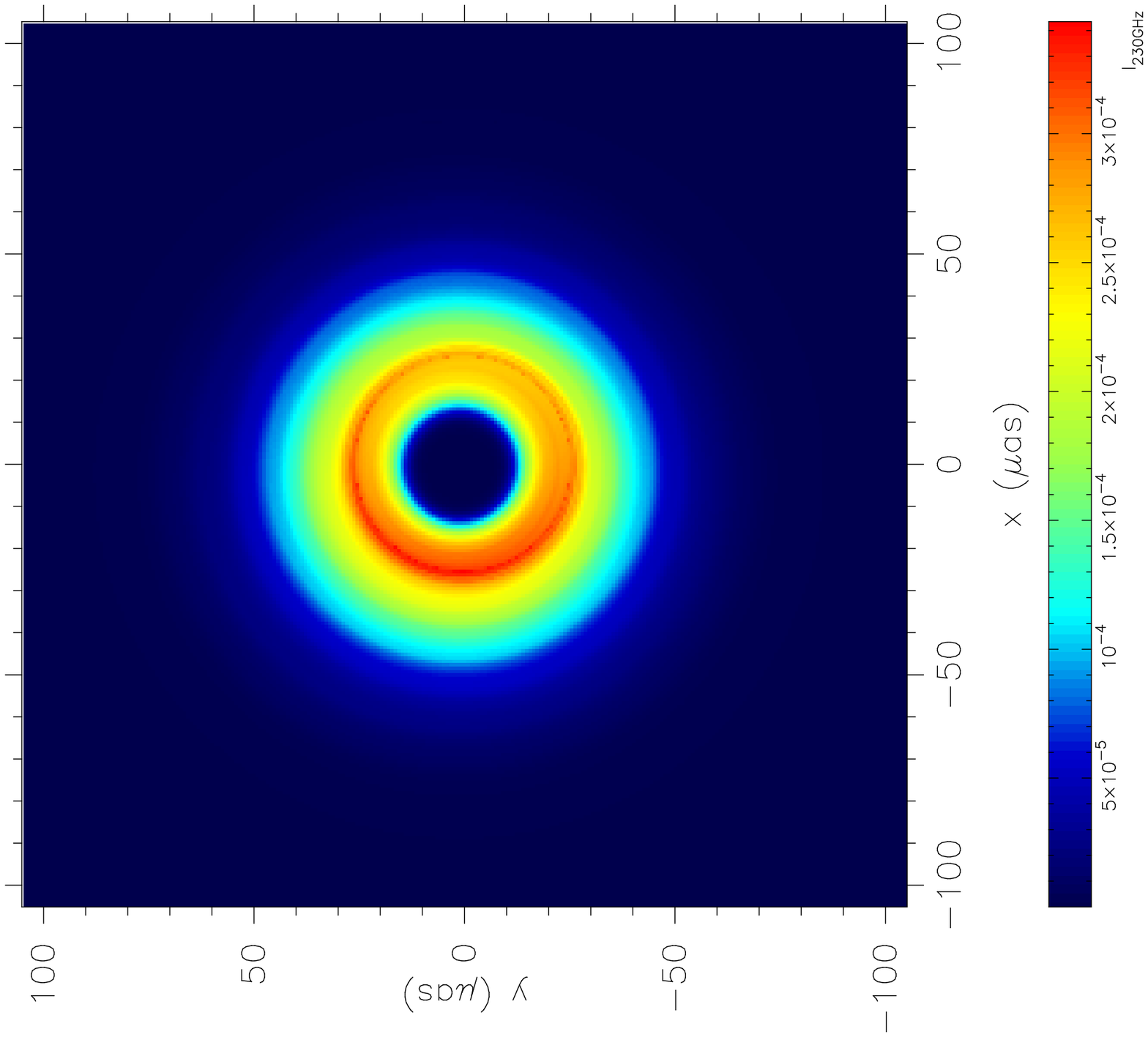}}
\put(90,400){\includegraphics{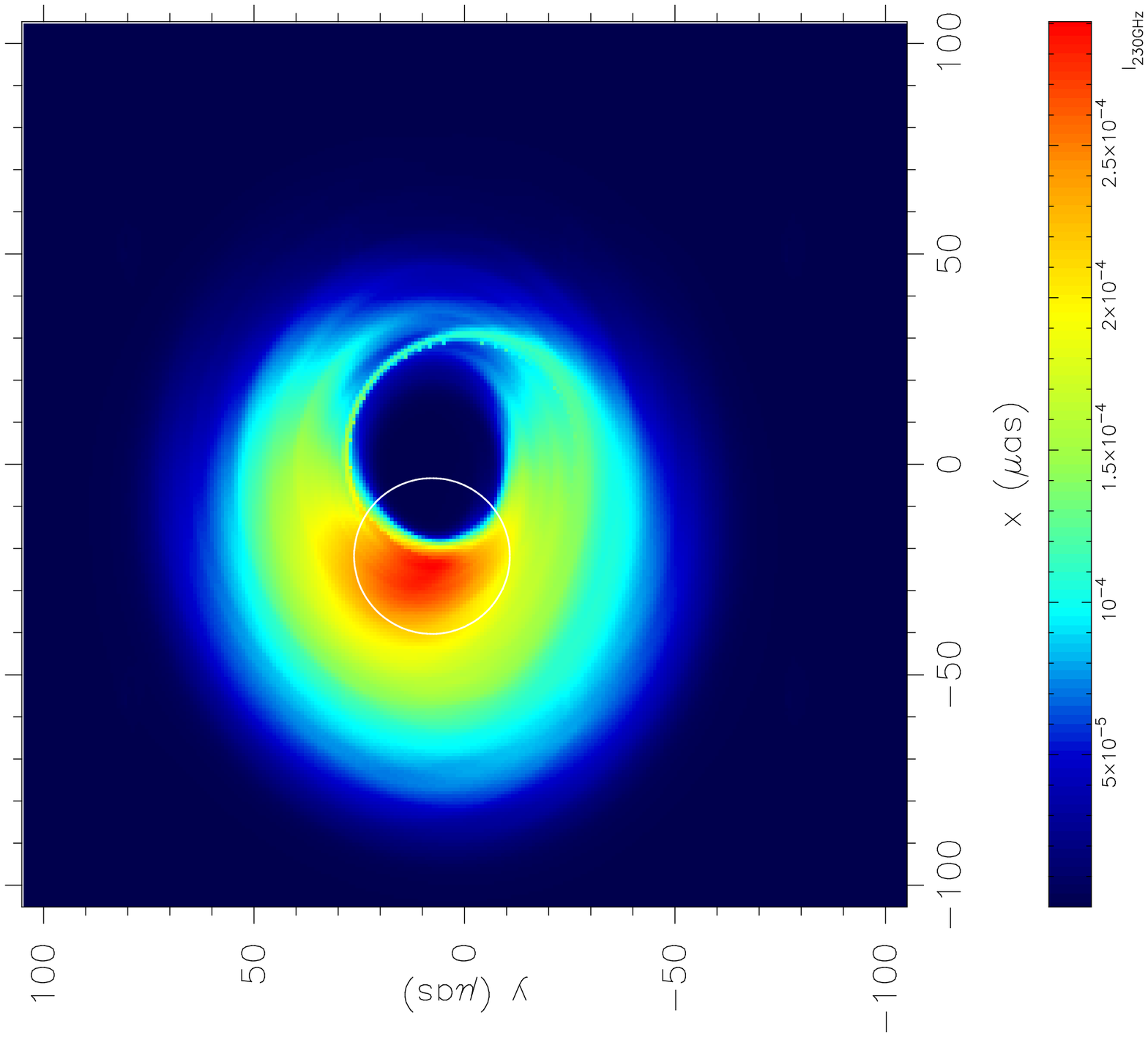}}

\put(-120,200){\includegraphics{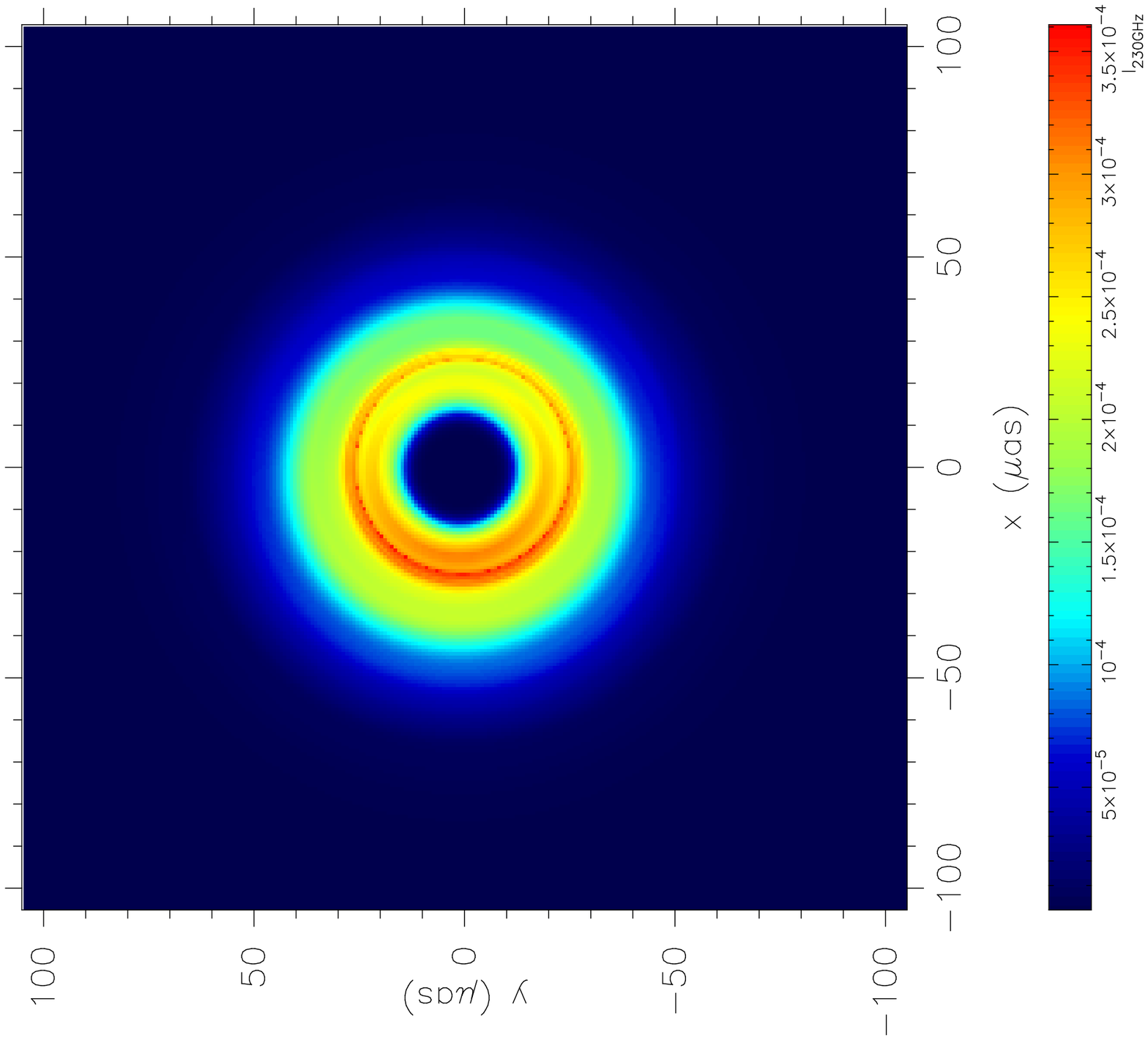}}
\put(90,200){\includegraphics{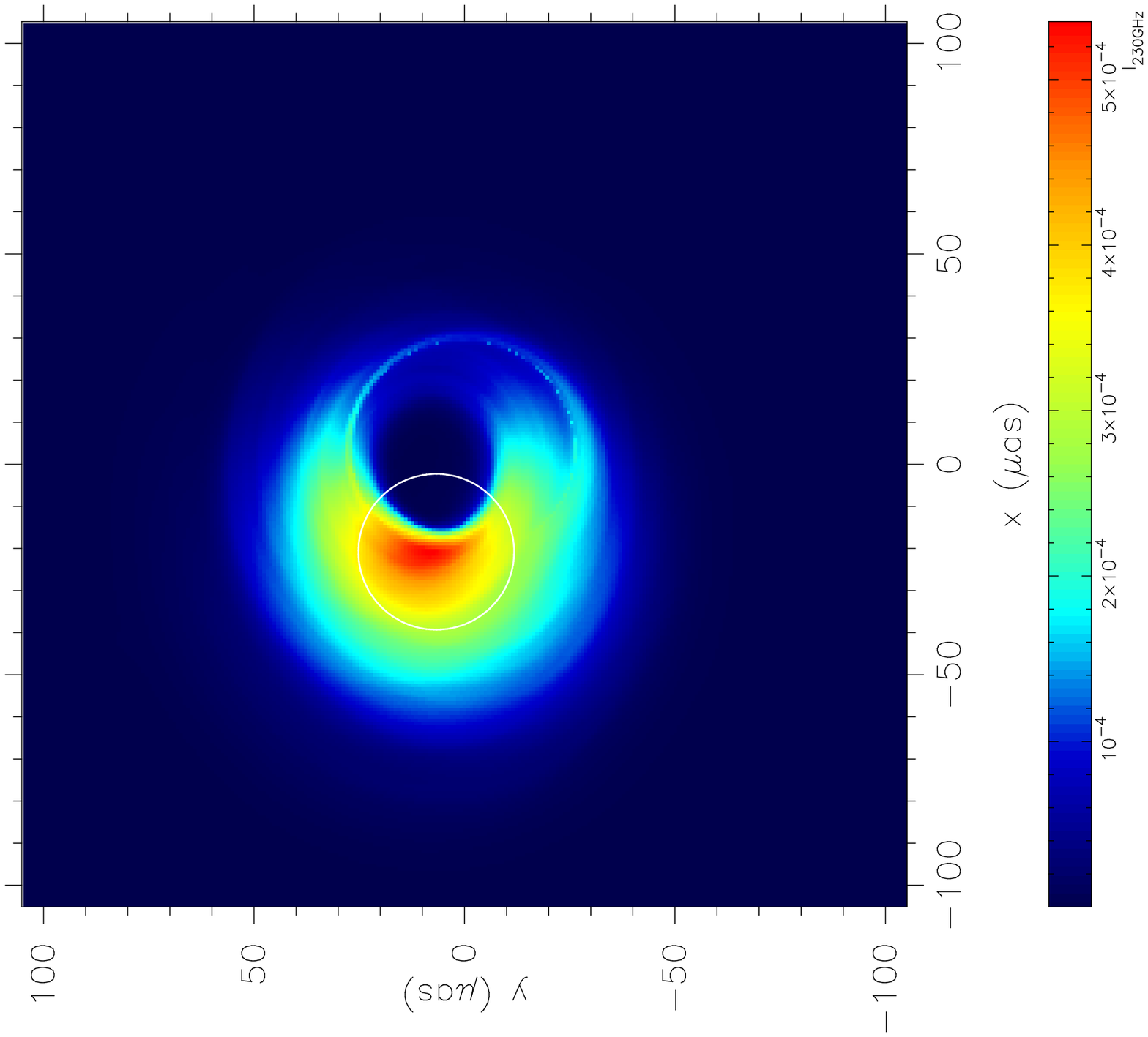}}

\end{picture}
\caption{
Images of the accretion flow at $230 \GHz$ for models with SEDs that are
consistent with observations of \sgra.  The images has been averaged
over time and over four separate realizations of each model. Intensities
are given in units of $\ergps {\rm pixelsize^{-2} Hz^{-1} sr^{-1}}$,
where the pixel size is $0.82 \muas$.  The images show
inner $40 \Rg$.  Left top panel shows model D with $\Trat=3$ at
$i=85\deg$. Left middle and bottom panels show high spin models E and F,
respectively, for $\Trat=10$ and $i=5\deg$. The right panels show models
A, B, and D for $\Trat=10$ and $i=45\deg$ in the upper, middle, and
lower panel respectively. The white circle marks FWHM=37 $\mu as$ of a
symmetric Gaussian brightness profile centered at the image centroid.
}
\label{fig:4}
\end{figure*}

We compute the 230 GHz intensity maps of our models using a ray tracing
code \citep{noble:2007} and we average them in the same manner as the
spectra. To estimate the size of the emitting region we calculate the
eigenvalues of the matrix formed by taking the second angular moments of
the image on the sky (i.e. the length of the ``principal axes'').  The
eigenvalues along the major ($\sigma_1$) and minor ($\sigma_2$) axis are
given in Table~\ref{tab:4}.  In Figure~\ref{fig:4} we show averaged
$230\GHz$ images for models with SEDs that are consistent with the data.  

The source size depends on the model parameters.  For $i>45\deg$, we
find a critical mass accretion rate $\dot{M} \approx 10^{-8} \MSUNYR$
(the exact value depends on $a_*$ and $i$) below which the size of the
emitting region decreases monotonically with increasing $\Trat$. Above
$\dot{M} \approx 10^{-8} \MSUNYR$, the size of the emitting region
increases with increasing $\Trat$. The increasing trend can be explained
by the appearance of the synchrotron photosphere at $230\GHz$  for
larger mass accretion rates at high inclination.  For $i=85\deg$ and
$\Trat=10$, at $230\GHz$ the black hole horizon is cloaked by the
photosphere and cannot be observed by VLBI (notice that this model is
ruled out for other reasons).  For a constant $\Trat$ and $i$ the size
always decreases with increasing $a_*$, because the emissivity of the
central regions increases with $a_*$.

The size of the emitting region for our best-bet model is consistent
with the observed FWHM $= 37 {\rm \muas}$ (inferred from VLBI data using
a two-parameter Gaussian model).  For $\Trat = 10$ the sizes of the
images are inconsistent with the VLBI measurement, except model D, which
is only marginally consistent.  Notice that this moment-based analysis
is crude.  It would be better to ``observe'' the model with the same
baselines used in gathering the VLBI data (this would add a new
parameter, the position angle).  Our analysis is particularly ill suited
to low $i$ models that are ring-like and therefore poorly fit by a
Gaussian model.

\subsection{Varying distance and mass}\label{sec:4.4}

In our discussion we have fixed the mass and distance of \sgra, but
these are uncertain to $\sim 5\%$.  How would changing these parameters
change our results?

First, consider how $\munit$ depends on $M$ and $D$.  Near the
submillimeter peak, $F_\nu \propto n_e B \lunit^3 D^{-2} \sim
\munit^{3/2} \lunit^{-1/2} \tunit^{-1} D^{-2} \sim (\munit/M)^{3/2}
D^{-2}$ (since $\lunit \sim M$ and $\tunit \sim M$) {\em if} the model
is optically thin (although there are usually optically thick lines of
sight through the model even if the mean optical depth at the
submillimeter peak is $< 1$).  We therefore expect that $\munit \sim
D^{4/3} M$.

Consider varying $M$ and $D$ in our best-bet model (model C; $\Trat=3$
and $i=85\deg$).  We find that $\Delta \munit / \munit \approx 15 \%$
when changing the distance from $8.0$kpc to $8.8$kpc if we fix $M=4.5
\times 10^{6} \MSUN$.  In particular, $\munit = 17.5 \times 10^{18},
18.9 \times 10^{18}, 20.6 \times 10^{18}$ for $D = 8.0, 8.4$ and $8.8$
kpc, respectively.  For $D = 8.4 \kpc$, $\munit=18.2 \times 10^{18},
18.9 \times 10^{18}$ and $20.6 \times 10^{18}$ for $M = 4.1, 4.5$ and
$4.9 \times 10^6 \MSUN$ respectively, which gives $\Delta \munit /
\munit \approx 10\%$ when changing black hole mass from $4.1-4.8 \times
10^{6} \MSUN$.  This is crudely consistent with our expectations based
on an optically thin source.

Finally we change the mass and distance simultaneously according to the
observational relation $M D^{-1.8}$=constant \citep{ghez:2008}.  For
$D=8.0 \kpc$ and $M=4.1 \times 10^{6} \MSUN$, $\munit=16.6 \times
10^{18}$ whereas for $D=8.8$ kpc and $M=4.8 \times 10^{6} \MSUN$,
$\munit=20.4 \times 10^{18}$.  We find that at $D=8.0 \kpc, M=4.1 \times
10^6 \MSUN$, $\alpha =-0.32$, $\log_{10} L_X=33.0$ ; at $D=8.4 \kpc,
M=4.5 \times 10^6 \MSUN$ $\alpha = -0.44$, $\log_{10} L_X=32.9$; at
$D=8.8 \kpc, M=4.8 \times 10^6 \MSUN$, $\alpha = -0.47$, $\log_{10}
L_X=32.7$.  The spectral slope and X-ray luminosity therefore vary $<
25\%$.

Our best-bet model remains consistent with the data, then, if we vary
with $\MBH$ and $D$ within the range permitted by observation.  Models
with $\Trat=10$ and $i=45\deg$ with BH spin $a_* = 0.97$ and $0.98$
become acceptable if $\MBH$ and $D$ are lowered, but only the model with
$a_* = 0.98$ would be (marginally) consistent with VLBI measurements of
the \sgra size.  In sum, $D$ and $\MBH$ are tightly constrained; varying
them within the narrow range of values permitted by observations does
not change the main conclusions of this work.

\section{Summary}\label{sec:5}

Under the assumption that the accretion flow at the galactic center is
optically thin, geometrically thick, and lightly magnetized, we have
presented constraints on $a_*$, $\Trat$, and $i$ for \sgra.  We find
that models with $\Trat = 3$ and $10$ describe the sub-mm spectral
observations ($\alpha$) better than models with $\Trat = 1$.  We find
that the model with $\Trat = 3$, black hole spin $a_* \approx 0.94$ and
the close to edge-on inclination angles is consistent with the broadband
SED observational data and the size of \sgra\ measured by VLBI. In this
case the silhouette of the black hole is difficult to observe because
Doppler boosting of the disk emission places almost all the emission on
one side of the black hole.

If, on the other hand, the electrons are heated relatively inefficiently
($\Trat=10$) then models with $a_* = 0.97, 0.98$ observed at $i=5\deg$,
or $a_* = 0.5, 0.75, 0.88, 0.94$ observed at $i=45\deg$ are consistent
with the observed SED.  The sizes of the emitting regions in these
models, however, seem to be inconsistent with the VLBI measurements,
except again at $a_* = 0.94$.

Our best-bet estimate of the black hole spin ($a_* = 0.94$) disagrees
with \citet{broderick:2008} (following \citealt{yuan:2009}) who found
$a_* = 0^{+0.4}$ and $i=90\deg_{-50\deg}$ ($2\sigma$ errors) based on a
careful analysis of images of RIAF models.  The discrepancy may be a
consequence of different emissivity (ours is based on Leung et al. 2009,
in preparation), and different underlying models for the run of
temperature, density profile, magnetic field strength, and geometry of
the flow.  We also do not include non-thermal emission as in
\citet{broderick:2008}.  The results for $\Trat=10$ and $i=45\deg$ at
low spin values agree with the previous study, but according to our
moment analysis these models are inconsistent with the VLBI data.  An
analysis of images at $\nu = 230\GHz$ that folds the models through the
VLBI observation process is needed to definitely exclude models based on
the VLBI data.  

Our $a_*$ and $i$ constraints are different from those presented in
\citet{noble:2007}, because here we allow $\Trat \ne 1$.  We do not find
a good fit to the observational data for single-temperature models which
were studied in the earlier work.  We confirm the trend that the
bolometric luminosity increases with the increasing black hole spin.

The models studied here differ in many respects from those
considered earlier by \citet{moscibrodzka:2007}.  The earlier models
were based on low angular momentum, nonrelativistic hydrodynamic models
for the accretion flow that extended over a wide range in radii.  The 
models described here are fully relativistic MHD models that extend over
a limited range in radius and use fully relativistic radiative
transfer.

There are still significant uncertainties in our models.  These
uncertainties fall into four categories: the unimportant; those which
may be important and be easily eliminated with a small additional
effort; those which may be important and require a major effort, but are
in principle straightforward to eliminate; and those which are serious
and require new physical understanding.

In the interests of full disclosure, our unimportant approximations are:
(1) we use a $\gamma$ law equation of state rather than a Synge-type
equation of state that would more accurately represent our
two-temperature relativistic gas.  Shiokawa et al. 2009 (in preparation)
show that the associated changes in spectra are small;  (2) for $\Trat
\gtrsim 10$ our cyclo-synchrotron emissivity and absorptivity are
imperfect because most of the emission comes from electrons with
$\Theta_e \sim 1$, where our approximate expression begins to break down
(Leung et al. 2009, in preparation) \footnote{For models with
$\Trat=1$ and $3$ the error in SED associated with our approximate
emissivity function is less than $1\%$ at all frequencies.  For
$\Trat = 10$ the errors are less than $10\%$ at $230 \GHz$ because
the emissivity-weighted mean temperature is lower and self-absorption is
important. The errors are taken from comparison of 
Leung et al. 2009 emissivity formula with directly integrated
cyclo-synchrotron harmonics for lowest
values of $\Theta_e$ found in our simulations.}; 
(3) we neglect bremsstrahlung, which is expected to be
important only far from the horizon; (4) we neglect double Compton
scattering.  The cross section for the double Compton process is $\sim
e^2/\hbar c$ (fine-structure constant) smaller than single Compton
scattering and can be neglected here (also $h\nu \ll m_e c^2$);  (5) we
neglect induced Compton scattering.  Induced Compton scattering is
important for $(k_B T_b/ m_e c^2) \tau_T \gtrsim 1$, where $T_b$ is the
brightness temperature.  In \sgra\ $T_b \approx 10^{10}$ K and $\tau_T
\ll 1$ so it is indeed negligible.  

The significant approximations that could be fixed with some additional
effort include: (1) our neglect of cooling.  It is straightforward to
run our GRMHD models with cooling, but then they are no longer
scale-free.  With cooling turned on we would need to fix $\munit$ by
evolving the GRMHD model at a trial $\munit$, calculating the spectrum,
and repeating until $F_\nu(230 \GHz) = 3.4\Jy$. (2) axisymmetry.  Three
dimensional (3D) models are available but far more expensive to evolve.
Use of a 3D models would permit us to evolve models with (3) a wider
range of radii, so that millimeter emission could be included and even
the sub-mm emission could be more accurately modeled.  We have not run
axisymmetric models with radially extended accretion flows because they
tend to develop pathologies (strong, radially extended magnetic
filaments).  (4) our neglect of nonthermal electrons.  These could be
readily included using a phenomenological prescription for the shape and
amplitude of the nonthermal portion of the electron distribution
function (\citealt{ozel:2000}, \citealt{yuan:2003},
\citealt{chan:2009}),  (5) our use of the steady-state approximation in
calculating SEDs and images.  This would require time-dependent
radiative transfer, which is straightforward in principle but
computationally expensive.

Other approximations can be fixed only with significant additional
effort: (1) our neglect of pair production.  This would require a model
for the radiation field near the pair production threshold.  Our
preliminary estimates suggest that in many of our models pair production
is substantial.  One advantage of incorporating pair production is that
it might permit us to eliminate our numerical floor and therefore more
accurately evolve the low density funnel region; (2) our treatment of
thermal energy in the funnel.  \harm\ tends to produce high
temperatures in the tenuous funnel plasma, some of which are clearly
numerical artifacts caused by application of the density floor and
other, more subtle, numerical issues associated with the small ratio of
thermal energy density to other energy densities; (3) our simplistic,
two-temperature thermal model for the plasma.  This includes our neglect
of conduction and anisotropy of the plasma.  

New physical understanding would be required to predictively model (1)
nonthermal parts of the distribution function and (2) the initial
magnetic field configuration.  Nonthermal particles can of course be
included in a phenomenological prescription, but the particle injection
and acceleration processes are still not fully understand.  As we have
already mentioned, prior work shows that the GRMHD models depend
nontrivially on the initial field configuration.  We have adopted a
simple, numerically appealing initial configuration, but the long-term
evolution of the large-scale field is ill understood.

Finally, notice that there are observational constraints from
polarization data and from light curves (statistically, the one and
two-point statistics of the light curves at each frequency, and the
cross correlations between different frequencies).  Treating the
polarization data requires accurate emissivities and absorptivities, as
well as models that extend well past the radius where $\Theta_e = 1$,
which is where most of the intrinsic Faraday rotation occurs.  The
light curves require full, time-dependent radiative transfer, since the
dynamical time is comparable to the light crossing time.

\acknowledgments

This work was supported by the National Science Foundation under grants
AST 00-93091, PHY 02-05155, and AST 07-09246, through TeraGrid resources
provided by NCSA and TACC, and by a Richard and Margaret Romano
Professorial scholarship, a Sony faculty fellowship, and a University
Scholar appointment to CFG.  The authors are grateful to Stu Shapiro,
Fred Lamb, Dan Marrone, Shep Doeleman, and Vincent Fish for discussion
and comments.

\clearpage

\begin{deluxetable}{ccccccccc}
\tabletypesize{\scriptsize}

\tablecaption{Summary of MHD and SED for models with $\Trat=1$\label{tab:1}}
\tablewidth{0pt}
\tablehead{

\colhead{run} & \colhead{$a_*$} & \colhead{$i$} &
\colhead{$\<\dot{M}\>$} & \colhead{$\alpha$} & \colhead{$\log_{10} L_X$ } & \colhead{$\eta$}&
\colhead{$\eta_{TD}$}&\colhead{consistent} \\ 
\colhead{} & \colhead{} & \colhead{[deg]} &
\colhead{$[10^{-9} \MSUNYR]$} & \colhead{} & \colhead{$[\ergps]$} & & & \colhead{w/ obs.?}
}

\startdata 
  &        &            5 & 3.97 & -2.04&30.9&$8.9 \times 10^{-3}$& &NO  \\ 
A & 0.5    &            45& 3.55 & -1.78&30.7&$8.0 \times 10^{-3}$& 0.0821&NO  \\ 
  &        &            85& 3.81 & -1.21&31.3&$8.6 \times 10^{-3}$& &NO  \\ 
\\
  &        &            5 & 1.6  & -1.94&31.1&$2.4 \times 10^{-2}$& &NO \\
B & 0.75   &            45& 1.39 & -1.69&31.0&$2.1 \times 10^{-2}$& 0.112&NO \\
  &        &            85& 1.5  & -1.12&31.6&$2.2 \times 10^{-2}$& &NO \\
\\
  &        &            5 & 0.94 & -1.92&31.3&$4.8 \times 10^{-2}$& &NO \\
C & 0.875  &            45& 0.87 & -1.65&31.3&$4.3 \times 10^{-2}$& 0.145& NO \\
  &        &            85& 0.91 & -1.07&32.3&$4.5 \times 10^{-2}$& & NO \\
\\
  &        &            5 & 0.86 & -1.67&31.9&$7.1 \times 10^{-2}$& & NO \\
D & 0.94   &            45& 0.78 & -1.41&32.1&$6.4 \times 10^{-2}$& 0.179& NO \\
  &        &            85& 0.80 & -0.87&33.1&$6.6 \times 10^{-2}$& & NO \\
\\
  &        &            5 & 0.85 & -1.51&32.3&0.11& &NO \\
E & 0.97   &            45& 0.78 & -1.25&32.7&0.1 & 0.213&NO \\
  &        &            85& 0.80 & -0.68&33.8&0.1 & &NO \\
\\
  &        &            5 & 0.64 & -1.51&32.9&0.19 & &NO \\
F &0.98   5&            45& 0.58 & -1.25&33.1&0.17 & 0.245&NO \\
  &        &            85& 0.59 & -0.69&34.2&0.18 & &NO \\
\enddata
\tablecomments{
The columns from left to right are: run ID, dimensionless spin of the black
hole, inclination angle of the observer with respect to the black hole spin
axis, averaged rest mass accretion rate, $\alpha$ spectral slope between
230-690 \GHz ($F\sim\nu^{\alpha}$), and luminosity in the X-rays (at $\nu \sim
10^{18}$ Hz), the radiative efficiency $\eta=L_{\rm BOL} / \dot{M} c^2$, the 
thin disk efficiency for the same $a_*$, and whether the model is consistent with
the data.}
\end{deluxetable}

\begin{deluxetable}{ccccccccc}
\tabletypesize{\scriptsize}

\tablecaption{Summary of MHD and SED for models with $\Trat=3$\label{tab:2}}
\tablewidth{0pt}
\tablehead{

\colhead{run} & \colhead{$a_*$} &  \colhead{$i$} &
\colhead{$\<\dot{M}\>$} & \colhead{$\alpha$} & \colhead{$\log_{10} L_X$ } & \colhead{$\eta$}&
\colhead{$\eta_{TD}$}&\colhead{consistent} \\ 
\colhead{} & \colhead{} & \colhead{[deg]} &
\colhead{$[10^{-9} \MSUNYR]$} & \colhead{} & \colhead{$[\ergps]$} & & & \colhead{w/ obs.?}
}

\startdata
  &        &            5 & 10.0 & -2.09& 30.5 &$3.4 \times 10^{-3}$&      &NO\\ 
A & 0.5    &            45& 8.8  & -1.72& 32.4 &$3.1 \times 10^{-3}$&0.0821&NO\\ 
  &        &            85& 10.7 & -0.75& 31.3 &$3.6 \times 10^{-3}$&      &NO\\ 
\\
  &        &            5 & 3.9 & -1.99& 31.0 &$9.5 \times 10^{-3}$&       &NO\\
B & 0.75   &            45& 3.5 & -1.59& 31.0 &$8.5 \times 10^{-3}$&0.112  &NO\\
  &        &            85& 4.1 & -0.67& 31.7 &$9.9 \times 10^{-3}$&       &NO\\
\\
  &        &            5 &  2.28& -1.97& 31.1 &$2.0 \times 10^{-2}$&     &NO\\
C & 0.875  &            45&  2.07& -1.54& 31.2 &$1.8 \times 10^{-2}$&0.145&NO\\
  &        &            85&  2.43& -0.56& 32.3 &$2.1 \times 10^{-2}$&     &NO\\
\\
  &        &            5 & 1.9 & -1.68& 31.5 & $3.5 \times 10^{-2}$&     &NO\\
D & 0.94   &            45& 1.7 & -1.27& 31.8 & $3.1 \times 10^{-2}$&0.179&NO\\
  &        &            85& 1.86 & -0.44& 32.9& $3.4 \times 10^{-2}$&     &YES\\
\\
  &        &            5 & 1.86 & -1.37& 32.2& $6.0 \times 10^{-2}$&     &NO\\
E & 0.97   &            45& 1.68 & -1.01& 32.5& $5.4 \times 10^{-2}$&0.213&NO\\
  &        &            85& 1.83 & -0.21& 33.8& $5.9 \times 10^{-2}$&     &NO\\
\\
  &        &            5 & 1.4  & -1.48& 32.6& 0.11&     &NO\\
F &0.98    &            45& 1.23 & -1.13& 32.9& 0.10&0.245&NO\\
  &        &            85& 1.29 & -0.26& 34.3& 0.10&     &NO\\
\enddata
\tablecomments{Columns same as in Table~\ref{tab:1}.}
\end{deluxetable}

\newpage

\begin{deluxetable}{ccccccccc}
\tabletypesize{\scriptsize}

\tablecaption{Summary of MHD and SED for models with $\Trat$=10.\label{tab:3}}
\tablewidth{0pt}
\tablehead{

\colhead{run} & \colhead{$a_*$} & \colhead{$i$} &
\colhead{$\<\dot{M}\>$} & \colhead{$\alpha$} & \colhead{$\log_{10} L_X$ } & \colhead{$\eta$}&
\colhead{$\eta_{TD}$}&\colhead{consistent} \\ 
\colhead{} & \colhead{} & \colhead{[deg]} &
\colhead{$[10^{-9} \MSUNYR]$} & \colhead{} & \colhead{$[\ergps]$} & & &\colhead{w/ obs.?}
}

\startdata
  &        &            5 & 90.8 & -1.37& 30.1 &$5.4 \times 10^{-4}$&       & NO\\ 
A & 0.5    &            45& 117.1& -0.2 & 31.4 &$6.7 \times 10^{-4}$& 0.0821&YES\\ 
  &        &            85& 369.0& 1.38 & 33.5 &$1.6 \times 10^{-3}$&       &NO\\ 
\\
  &        &            5 & 38.3 & -1.05& 30.0 &$1.6 \times 10^{-3}$&      &NO\\
B & 0.75   &            45& 50.2 & 0.04 & 31.8 &$2.0 \times 10^{-3}$& 0.112&YES\\
  &        &            85& 190.6& 1.49 & 34.6 &$6.9 \times 10^{-3}$&      &NO\\
\\
  &        &            5 & 19.5 & -1.15& 31.4 &$5.1 \times 10^{-3}$&      &NO\\
C & 0.875  &            45& 23.6 & -0.07& 32.0 &$6.2 \times 10^{-3}$& 0.145&YES\\
  &        &            85& 41.4 & 1.19 & 34.2 &$1.7 \times 10^{-2}$&      &NO\\
\\
  &        &            5 & 13.7 & -0.93 & 31.8 &$1.1 \times 10^{-2}$&      &NO\\
D & 0.94   &            45& 15.2 & -0.05 & 32.3 &$1.1 \times 10^{-2}$& 0.179&YES\\
  &        &            85& 31.2 & 1.17  & 34.4 &$2.5 \times 10^{-2}$&      &NO\\
\\
  &        &            5 & 13.6 & -0.40 & 32.5 &$2.6 \times 10^{-2}$&      &YES\\
E & 0.97   &            45& 14.3 & 0.2   & 33.1 &$2.8 \times 10^{-2}$& 0.213&NO\\
  &        &            85& 26.5 & 1.19  & 35.2 &$5.1 \times 10^{-2}$&      &NO\\
\\
  &        &            5 & 9.07 & -0.6  & 32.7 &$5.3 \times 10^{-2}$&      &YES\\
F &0.98    &            45& 9.16 & -0.08 & 33.2 &$5.2 \times 10^{-2}$& 0.245&NO\\
  &        &            85& 15.5 & 1.12  & 35.4 &$9.0 \times 10^{-2}$&      &NO\\
\enddata
\tablecomments{Columns same as in Table~\ref{tab:1}.}
\end{deluxetable}

\newpage

\begin{deluxetable}{ccccccccccc}
\tabletypesize{\scriptsize}
\tablecaption{The size of the emitting region at $230 \GHz$.\label{tab:4}}
\tablewidth{0pt}
\tablehead{
   \colhead{}& \colhead{} &\colhead{}&\multicolumn{2}{c}{$\Trat=1$}
   & &  \multicolumn{2}{c}{$\Trat=3$}&& \multicolumn{2}{c}{$\Trat=10$} \\
   \colhead{run}& \colhead{$a_*$} &\colhead{$i$}& \colhead{$\sigma_1$ }&
   \colhead{$\sigma_2$} & &\colhead{$\sigma_1$}& \colhead{$\sigma_2$}& &
   \colhead{$\sigma_1$} & \colhead{$\sigma_2$}
}
\startdata
       &       & 5   & 41.7 & 41.6 &   & 38.1 & 38.0&    & 35.9& 35.9\\
     A & 0.5   & 45  & 31.2 & 28.4 &   & 28.9 & 25.9&    & 32.7& 31.6\\
       &       & 85  & 23.1 & 20.7 &   & 23.3 & 21.9&    & 37.6& 33.9\\
\\
       &       & 5   & 38.8  & 38.7&   & 35.4 & 35.3&    & 34.6& 34.6\\
     B &0.75   & 45  & 28.6  & 25.9&   & 26.7 & 23.7&    & 31.2& 30.2\\
       &       & 85  & 20.7  & 19.4&   & 20.8 & 20.4&    & 36.9& 32.3\\
\\
       &       & 5   & 39.3  & 39.2&   & 35.2 & 35.0&    & 31.9& 31.8\\
     C &0.875  & 45  & 29.3  & 26.7&   & 26.4 & 23.4&    & 29.3&28.0 \\
       &       & 85  & 20.6  & 19.4&   & 20.4 & 19.7&    & 30.6& 30.1\\
\\
       &       & 5   & 37.2  & 37.1&   & 32.2 & 32.1&    & 28.0& 28.0\\
     D &0.94   & 45  & 27.4  & 25.3&   & 24.0 & 21.5&    & 24.7& 23.9\\
       &       & 85  & 19.2  & 18.4&   & 18.5 & 17.2&    & 27.1& 26.4\\
\\
       &       & 5   & 37.0  & 36.9&   & 31.2 & 31.1&    & 26.5& 26.4 \\
     E &0.97   & 45  & 27.4  & 25.7&   & 23.4 & 21.6&    & 23.6& 23.4\\
       &       & 84  & 19.6  & 18.2&   & 18.1 & 17.3&    & 25.8& 24.6\\
\\
       &        & 5  & 36.5  & 36.4&   & 31.3 & 31.2&    & 26.1& 26.1\\
     F &0.98    & 45 & 27.2  & 25.2&   & 23.4 & 21.2&    & 22.8& 21.7\\
       &        & 85 & 18.8  & 18.4&   & 16.5 & 17.6&    & 23.8& 22.6 \\
\enddata
\tablecomments{
The columns from left to right are: run ID, dimensionless spin of the black
hole, inclination angle $i$, size of the emitting region in
terms of standard deviation in the major ($\sigma_1$) and minor
($\sigma_2$) axis for $\Trat=1$ (col. 4 and 5), $\Trat=3$ (col. 6 and 7),
and $\Trat =10$ (col. 6 and 9) in units of ${\rm \mu as}$.  For a
Gaussian model, the VLBI data require FWHM$ = 37^{+16}_{-10} \muas$
\citep{doeleman:2008}, or $\sigma = 16^{+6.8}_{-4.2}\muas$.}
\end{deluxetable}

\newpage

\newpage

\newpage

\newpage

\newpage

\end{document}